\documentclass[floats,floatfix,showpacs,amssymb,prd,twocolumn,superscriptaddress,nofootinbib,nolongbibliography,reprint,preprintnumbers]{revtex4-1}

\usepackage{amssymb,amsmath,verbatim,mathtools,needspace,enumitem,etoolbox,graphicx,physics,microtype,afterpage,xspace,tabularx,lmodern,multirow,bm}
\usepackage{textcomp, gensymb}

\usepackage{braket}

\usepackage{relsize}

\usepackage{dcolumn}

\usepackage[normalem]{ulem}
\usepackage{placeins}
\usepackage{comment}
\usepackage[dvipsnames, usenames]{xcolor}
\definecolor{linkcolor}{rgb}{0.0,0.3,0.5}
\usepackage[unicode, colorlinks=true, linkcolor=linkcolor, citecolor=linkcolor, filecolor=linkcolor, urlcolor=linkcolor, linktocpage, breaklinks]{hyperref}
\usepackage[all]{hypcap}
\usepackage[T1]{fontenc}
\usepackage[utf8]{inputenc}
\usepackage{mathrsfs}
\usepackage[usenames,dvipsnames]{xcolor}
\hypersetup{colorlinks=true,citecolor=romared,linkcolor=romared,urlcolor=romared}

\definecolor{romared}{RGB}{142,0,28}

\newcommand{\be}{\begin{equation}}
\newcommand{\ee}{\end{equation}}

\def\be{\begin{equation}}
\def\ee{\end{equation}}
\newcommand{\beq}{\begin{eqnarray}}
\newcommand{\eeq}{\end{eqnarray}}

\usepackage{aas_macros}
\usepackage{makecell}
\usepackage{soul}
\usepackage[nolist,nohyperlinks]{acronym}
\usepackage{orcidlink}
\usepackage{lipsum}

\acrodef{LSC}[LSC]{LIGO Scientific Collaboration}
\acrodef{BH}{black hole}
\acrodef{NS}{neutron star}
\acrodef{PN}{Post-Newtonian}
\acrodef{BBH}{binary black-hole}
\acrodef{BNS}{binary neutron-star}
\acrodef{NSBH}{neutron-star black-hole}
\acrodef{NR}{numerical relativity}
\acrodef{GW}{gravitational wave}
\acrodef{PSD}{power spectral density}
\acrodef{aLIGO}{Advanced Laser interferometer Gravitational-Wave Observatory}
\acrodef{AZDHP}{aLIGO zero detuned high power density}
\acrodef{GR}{general relativity}
\acrodef{PE}{parameter estimation}
\acrodef{LAL}{LIGO algorithm library}
\acrodef{TPI}{tensor-product interpolant}
\acrodef{SVD}{singular value decomposition}
\acrodef{SNR}{signal-to-noise ratio}
\acrodef{ODE}{ordinary differential equation}
\acrodef{PDE}{partial differential equation}
\acrodef{ROM}{reduced order model}
\acrodef{QNM}{quasi-normal mode}
\acrodef{IMR}{inspiral-merger-ringdown}
\acrodef{LVK}{LIGO-Virgo-KAGRA}
\acrodef{SXS}{Simulating eXtreme Spacetimes}

\newcommand{\jhu}{\affiliation{William H. Miller III Department of Physics and Astronomy, Johns Hopkins University, 3400 North Charles Street, Baltimore, Maryland, 21218, USA}}
\newcommand{\gssi}{\affiliation{Gran Sasso Science Institute (GSSI), I-67100 L’Aquila, Italy}}
\newcommand{\infngssi}{\affiliation{INFN, Laboratori Nazionali del Gran Sasso, I-67100 Assergi, Italy}}
\newcommand{\sissa}{\affiliation{SISSA, Via Bonomea 265, 34136 Trieste, Italy and INFN Sezione di Trieste}}
\newcommand{\ifpu}{\affiliation{IFPU - Institute for Fundamental Physics of the Universe, Via Beirut 2, 34014 Trieste, Italy}}

\graphicspath{{../Plots/}}

\newcommand{\orcid}[1]{\href{https://orcid.org/#1}{\includegraphics[width=10pt]{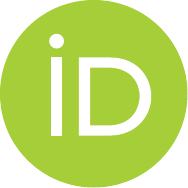}}}
\newcommand{\ben}{\begin{enumerate}}
\newcommand{\een}{\end{enumerate}}

\def\be{\begin{equation}}
\def\ee{\end{equation}}
\def\beq{\begin{eqnarray}}
\def\eeq{\end{eqnarray}}

\def\araa{\rm{ARA\&A}}

\def\apjl{\rm{ApJ}}
\def\apjs{\rm{ApJS}}
\def\aap{\rm{A\&A}}

\def\mnras{\rm{MNRAS}}
\def\prd{\rm{PRD}}

\graphicspath{{./Plots/}}

\allowdisplaybreaks

\begin{document}

\pagenumbering{arabic}

\title{Nonlinear quasinormal mode detectability with next-generation gravitational wave detectors}

\author{Sophia Yi \orcid{0000-0002-9104-1734}}
\email{syi24@jh.edu}
\jhu
\author{Adrien Kuntz \orcid{0000-0002-4803-2998}}
\sissa
\ifpu
\author{Enrico Barausse \orcid{0000-0001-6499-6263}}
\sissa
\ifpu
\author{Emanuele Berti \orcid{0000-0003-0751-5130}}
\jhu
\author{Mark Ho-Yeuk Cheung \orcid{0000-0002-7767-3428}}
\jhu
\author{Konstantinos Kritos \orcid{0000-0002-0212-3472}}
\jhu
\author{Andrea Maselli \orcid{0000-0001-8515-8525}}
\gssi
\infngssi

\pacs{}
\date{\today}

\begin{abstract}
In the aftermath of a binary black hole merger event, the gravitational wave signal emitted by the remnant black hole is modeled as a superposition of damped sinusoids known as quasinormal modes. While the dominant quasinormal modes originating from linear black hole perturbation 
theory have been studied extensively in this postmerger ``ringdown'' phase, more accurate models of ringdown radiation include the nonlinear modes 
arising from higher-order perturbations of the remnant black hole spacetime. We explore the detectability of quadratic quasinormal modes with both ground- and space-based next-generation detectors. We quantify how predictions of the quadratic mode detectability depend on the quasinormal mode starting times. We then calculate the signal-to-noise ratio of quadratic modes for several detectors and binary black hole populations, focusing on the ($220\times220$) mode -- i.e., on the quadratic term sourced by the square of the linear $(220)$ mode. For the events with the loudest quadratic mode signal-to-noise ratios, we additionally compute statistical errors on the mode parameters in order to further ascertain the distinguishability of the quadratic mode from the linear quasinormal modes. The astrophysical models used in this paper suggest that while the quadratic mode may be detectable in at most a few events with ground-based detectors, the prospects for detection with the Laser Interferometer Space Antenna (LISA) are more optimistic.
\end{abstract}

\preprint{ET-0081A-24}

\maketitle

\section{\label{sec:intro}Introduction}

The postmerger signal of a binary black hole (BH) merger can be accurately modeled by a superposition of the oscillation modes of a single perturbed BH, known as quasinormal modes (QNM)~\cite{Kokkotas:1999bd,Berti:2009kk,Berti:2018vdi}.
The full merger waveform can be computed in numerical relativity, and the predictions of linear perturbation theory around a fixed BH background provide an excellent fit of the ``ringdown'' phase of the waveform at late times~\cite{Buonanno:2006ui, Berti:2007dg}.

In linear perturbation theory, the time-domain waveform in the ringdown phase is a sum of damped sinusoids
with discrete frequencies labeled by the angular momentum numbers $(\ell m)$ of the spherical harmonics expansion and by an additional ``radial'' overtone number, $n$. These frequencies can be computed within linear perturbation theory, and they are intrinsic properties of the BH, depending only on its mass and spin. On the other hand, the amplitudes of the QNMs excited in specific astrophysical scenarios depend on the initial conditions of the perturbation. For a binary BH merger, these amplitudes depend on the parameters of the progenitor binary, and they are usually found by fitting either the data or the numerical simulations (see, e.g.,~\cite{Buonanno:2006ui,Berti:2007dg,Giesler:2019uxc,Cook:2020otn,MaganaZertuche:2021syq,Isi:2021iql,Baibhav:2023clw,Cheung:2023vki,Takahashi:2023tkb,Qiu:2023lwo}). The simplicity of the QNM spectrum in linear perturbation theory has triggered an observational research program to test the Kerr nature of the merger remnant by performing ``BH spectroscopy''~\cite{Detweiler:1980gk, Dreyer:2003bv, Berti:2005ys, Berti:2016lat, Bhagwat:2016ntk, Yang:2017zxs, Baibhav:2017jhs, Brito:2018rfr, Baibhav:2018rfk, Bhagwat:2019dtm, Maselli:2019mjd, Bhagwat:2019dtm, CalderonBustillo:2020rmh, Capano:2020dix, JimenezForteza:2020cve, Ota:2021ypb, Maselli:2023khq, Baibhav:2023clw}. A measurement of more than one of the complex QNM frequencies could test the so-called ``no-hair theorem'' of general relativity (GR), because all of the frequencies depend only on the mass and spin of the remnant BH.
The large number of events and the large signal-to-noise ratios (SNRs) expected for next-generation (XG) gravitational wave interferometers, such as the Einstein Telescope (ET)~\cite{Punturo:2010zz} in Europe, Cosmic Explorer (CE)~\cite{2015PhRvD..91h2001D} in the US, or LISA~\cite{LISA:2017pwj,Colpi:2024xhw}, 
suggest that precision BH spectroscopy may become routine in the near future.

However, the idea that linear perturbation theory is sufficient to describe the ringdown phase has recently been challenged. Einstein's theory of GR is well known to be nonlinear, and there could be an imprint of these nonlinearities in the QNMs. On the theoretical side, second-order perturbation theory of BH backgrounds is an active field of research (see e.g.~\cite{Gleiser:1995gx,Nicasio:1998aj,Gleiser:1998rw,Campanelli:1998jv,Zlochower:2003yh,Brizuela:2009qd,Brizuela:2007zza,Brizuela:2006ne,Loutrel:2020wbw,Ripley:2020xby,Sberna:2021eui,Wardell:2021fyy,Spiers:2023mor,Kehagias:2023ctr,Redondo-Yuste:2023seq,Perrone:2023jzq,Bucciotti:2023ets,Zhu:2024rej,Ma:2024qcv,Spiers:2024src}). The  second-order perturbations of the metric around a Schwarzschild BH obey a very similar differential equation to the linear ones, the main difference being the presence of a term sourcing the perturbations in the Regge-Wheeler or Zerilli equations. The nonlinear source term depends on a product of two linear modes which can source nonlinear QNMs. 
This property implies that at second order there must be a set of  ``quadratic modes'' in the ringdown waveform, besides the usual linear QNMs~\cite{Ioka:2007ak,Nakano:2007cj,Okuzumi:2008ej,Pazos:2010xf,Lagos:2022otp}. The frequency of a nonlinear mode is just the sum or difference of the two linear QNM frequencies $(\ell_1 m_1n_1 \times \ell_2 m_2 n_2)$ entering the source term, a typical feature of nonlinear systems. Furthermore, several works~\cite{Ioka:2007ak,Nakano:2007cj,Lagos:2022otp,Bucciotti:2023ets,Redondo-Yuste:2023seq,Zhu:2024rej,Ma:2024qcv} highlighted that the amplitudes of the quadratic modes are also an intrinsic property of the BH, once the amplitudes of linear QNMs are fixed by the initial conditions. This makes the observation of quadratic modes a very interesting target to perform further tests of GR in the nonlinear regime.

On the numerical side, the in-depth study of numerical relativity simulations has confirmed the existence of quadratic QNMs and provided fits to their amplitudes as a function of the progenitor parameters~\cite{London:2014cma,Mitman:2022qdl,Cheung:2022rbm,Ma:2022wpv,Redondo-Yuste:2023seq,Cheung:2023vki,Takahashi:2023tkb,Qiu:2023lwo}.  These works identify the $(220 \times 220)$ mode, generated by the self-interaction of the $(220)$ mode, as the dominant quadratic mode in the waveform.

From an observational perspective, however, the detectability of quadratic modes is still an open question.  Being generated at second-order in perturbation theory, the amplitude of these modes is expected to be smaller than the amplitude of the linear modes sourcing them, while their damping time is always shorter~\cite{Ioka:2007ak,Nakano:2007cj,Lagos:2022otp}.  These two properties combine to make the potential detection of quadratic QNMs in data a challenging task, probably out of reach for current interferometers. However, the prospects may be more optimistic in the future with the advent of XG detectors.  In this article we investigate the possibility of observing and measuring quadratic QNMs in XG gravitational wave observatories, both on the ground and in space.

The paper is organized as follows.
In Sec.~\ref{sec:datanalysis} we describe our ringdown waveform model and data analysis framework.
In Sec.~\ref{sec:catalogs} we describe the astrophysical catalogs used to estimate the quadratic mode detection rates.
In Sec.~\ref{sec:detectability} we discuss the detectability of the quadratic modes, and in Sec.~\ref{sec:measurability} the measurability of their parameters.
In Sec.~\ref{sec:conclusions} we present our main conclusions and some directions for future work.
To improve readability, some technical material (on the effect of the QNM starting time, the calculation of binary BH merger rates in clusters, and the calculation of SNRs and parameter estimation errors) is presented in the appendices.
Throughout the paper we use geometrical units ($G=c=1$).

\section{Waveform models and data analysis framework}
\label{sec:datanalysis}

\subsection{Ringdown waveform template}\label{sec:templates}

We assume that the plus and cross time-domain polarizations of the gravitational waveform in the ringdown stage can be written as~\cite{Berti:2007zu, Berti:2007fi}
\begin{align}
h_+(t)&=\sum_{\ell m n} \mathrm{Re} \left[ \Bar{\mathcal{A}}_{\ell m n} e^{i \phi_{\ell mn}} \hat{Y}^{\ell m}_+ e^{i(\omega_{\ell mn} + i/\tau_{\ell mn})t}\right], \label{eq:hp_timedomain}\\
    h_\times(t)&=\sum_{\ell m n} \mathrm{Im} \left[ \Bar{\mathcal{A}}_{\ell m n} e^{i \phi_{\ell mn}} \hat{Y}^{\ell m}_\times e^{i(\omega_{\ell mn} + i/\tau_{\ell mn})t}
\right]\ ,\label{eq:h_ringdown}
\end{align}
where $\phi_{\ell mn}$ is the QNM phase and $\omega_{\ell mn}$ and $\tau_{\ell mn}$ are the QNM frequencies and damping times, respectively, all specified by the angular indices $\ell m$ and by the overtone number $n$. 
Following Ref.~\cite{Maselli:2023khq} we have also introduced an effective amplitude $\Bar{\mathcal{A}}_{\ell mn}=M_f\mathcal{A}_{\ell mn}/r$, where $M_f$ is the redshifted mass of the remnant BH, $r$ is the luminosity distance, and $\mathcal{A}_{\ell m n}$ is the QNM amplitude.
In our analysis we neglect the contribution of the overtones (see e.g.~\cite{Baibhav:2023clw,Nee:2023osy}), and therefore from now on we generally omit the overtone index $n$ to simplify the notation.

The modes sum in Eqs.~\eqref{eq:hp_timedomain}-\eqref{eq:h_ringdown} 
includes: (i) the linear (22), (21), (33), and (44) 
QNMs and (ii) the loudest quadratic $(44)$ 
component, sourced by the square of the $(22)$ fundamental mode, with frequencies and damping 
times $\omega_{22 \times 22} = 2 \omega_{22}, \tau_{22 \times 22} = \tau_{22}/2$. We obtain numerical 
values for QNM frequencies from Ref.~\cite{Berti:2009kk} and mode amplitudes and phases from Ref.~\cite{Cheung:2023vki}, as we will discuss in more detail in Sec.~\ref{subsec:starting_time}.

The complex functions $\hat{Y}^{\ell m}_{+,\times}(\iota,\varphi)$ are defined by
\begin{equation}
\hat{Y}^{\ell m}_{+,\times}(\iota,\varphi)=e^{-im\varphi}\,Y^{\ell m}_{+,\times}(\iota,0)\ ,\label{eqn:yplus_ycrossh}
\end{equation}
where
\begin{align}
Y^{\ell m}_+(\iota,0)&=_{-2}Y^{\ell m}(\iota,0) + (-1)^\ell\, _{-2}Y^{\ell-m}(\iota,0)\ ,\\
 Y^{\ell m}_\times(\iota,0)&= _{-2}Y^{\ell m}(\iota,0) - (-1)^\ell\, _{-2}Y^{\ell-m}(\iota,0)\ ,
    \label{eqn:yplus_ycross}
\end{align}
and $_{-2}Y^{\ell m}(\iota,\varphi)=e^{im \varphi} {_{-2}Y^{\ell m}(\iota,0)}$ are the spin-weighted spherical harmonics.

We map the signal in the Fourier domain following the Flanagan-Hughes convention \cite{Flanagan:1997sx, Berti:2005ys}, such that the two polarizations read
\begin{align}
\Tilde{h}_+(f)&=\sum_{\ell m}\frac{\Bar{\mathcal{A}}_{\ell m}}{\sqrt{2}}\left[b_+ e^{i\phi_{\ell m}} \hat{Y}^{\ell m}_+ + b_- e^{-i\phi_{\ell m}} \hat{Y}^{\ell m*}_+ \right]\ , \label{eq:tildehp} \\
\Tilde{h}_\times(f)&=\sum_{\ell m}\frac{\Bar{\mathcal{A}}_{\ell m}}{i\sqrt{2}}\left[b_+ e^{i\phi_{\ell m}} \hat{Y}^{\ell m}_\times - b_- e^{-i\phi_{\ell m}} \hat{Y}^{\ell m*}_\times \right]\ ,\label{eq:tildeh}
\end{align}
where the superscript $^{*}$ denotes complex conjugation and $b_{\pm}$ are Breit-Wigner functions:
\begin{equation}
    b_{\pm}=\frac{1/\tau_{\ell m}}{\tau_{\ell m}^{-2}+(\omega \pm \omega_{\ell m})^2}\ ,\qquad \omega=2\pi f\ .
\end{equation}
The full ringdown waveform is then given by
\begin{equation}
    \Tilde{h}(f)=F_+ \Tilde{h}_+(f) + F_\times \Tilde{h}_\times(f)\ ,
    \label{eqn:waveform_full}
\end{equation}
where $F_{+,\times}$ are the detector pattern functions.  For simplicity, we average over the detector angles and over the source orientation. For L-shaped detectors we 
make use of the following identities: 
\begin{equation}
    \braket{F^2_{+,\times}}=\frac{1}{5}\quad \ ,\quad \braket{F_+F_\times}=0\ \, .\label{eq:antennaaverage}
\end{equation}
For LISA, Eqs.~\eqref{eq:antennaaverage} need to be multiplied by an additional factor of $3/2$ to take into account the triangular shape of the detector, as discussed in Ref.~\cite{Robson:2018ifk}.

The template \eqref{eqn:waveform_full} allows us to compute the SNR $\rho$ for a detector with noise power spectral density $S_n(f)$
\begin{equation}
    \rho^2= 4 \int_{f_{\mathrm{min}}}^{f_{\mathrm{max}}} \frac{\tilde{h}(f)\tilde{h}^*(f)}{S_n(f)} df\ ,\label{eqn:snr}
\end{equation}
which we use as a simple metric to determine the detectability of quadratic modes with both ground-based detectors and the space-based interferometer LISA~\cite{LISA:2017pwj,Colpi:2024xhw}. We consider a ground-based detector network of two L-shaped, aligned ET detectors with $15$\,km armlength \cite{Branchesi:2023mws} and a $40$\,km CE~\cite{Evans:2023euw}. The total SNR for the ET network is simply given by multiplying the single-detector SNR by a factor of $\sqrt{2}.$
We set $f_\mathrm{min}=3$\,Hz ($f_\mathrm{min}=10^{-5}$\,Hz) and $f_\mathrm{max}=5000$\,Hz ($f_\mathrm{max}=0.5$\,Hz) for ground (space) observations, respectively.

Note that since the waveform~\eqref{eqn:waveform_full} contains a combination of modes with different values of $(\ell m)$, the integral \eqref{eqn:snr} depends both on the SNR of the individual QNMs, $\rho_{\ell m}$, and on cross-products among them. However, due to the orthogonality of the spin-weighted spherical harmonics (see Appendix~\ref{app:orthogonality}) the total SNR of QNMs with different multipolar indices is simply given by the sum in quadrature of the individual $\rho_{\ell m}$'s, i.e.,
\begin{equation}
\rho=\sqrt{\rho^2_{\ell_1 m_1}+\rho^2_{\ell_2 m_2}+\ldots}\ .    
\end{equation}

\subsection{Fisher analysis}\label{sec:fisher_intro}

To assess the measurement accuracy of nonlinear modes by XG detectors we use a Fisher information matrix (FIM) approach~\citep{Vallisneri:2007ev}.  Given the strain $s(t)=h(t,\vec{\theta})+n(t)$, where $n(t)$ is the detector’s stationary noise and $h(t,\vec{\theta})$ the gravitational wave signal, the posterior probability of the waveform parameters $\vec{\theta}$ is given by
\begin{equation}
  p(\vec{\theta}\vert s)\propto p_0(\vec{\theta})
  e^{-\frac{1}{2}\Gamma_{ij}\Delta\theta_i\Delta\theta_j}\ ,
\end{equation}
where $p_0(\vec{\theta})$ corresponds to the prior distribution for $\vec{\theta}$, $\Delta\vec{\theta}=\vec{\theta}-\vec{\xi}$, and the vector $\vec{\xi}$ refers to the \textit{true} values of the parameters.
The FIM $\Gamma_{ij}$ is then defined as
\begin{equation}
    \Gamma_{ij}= \left( \frac{\partial h}{\partial \theta_i} \bigg|\frac{\partial h}{\partial \theta_j} \right)\ ,\label{eq:fisher}
\end{equation}
where we have introduced the inner product between two waveforms in the frequency domain,
\begin{equation}
    \left( h_1 | h_2 \right) = 2 \int^{f_{\mathrm{max}}}_{f_{\mathrm{min}}} \frac{\tilde{h}_1(f) \tilde{h}_2^* (f)+\tilde{h}_1^* (f)\tilde{h}_2(f)}{S_n(f)} df\ . \label{eq:inner_product}
\end{equation}
$h$ is given by Eq.~\eqref{eqn:waveform_full}, and Eq.~\eqref{eq:fisher} is evaluated at $\vec{\theta}=\vec{\xi}$.
Note that each mode contributes to $\vec{\theta}$ with four parameters $(\bar{\mathcal{A}}_{\ell m},\phi_{\ell m}, \omega_{\ell m},\tau_{\ell m})$,
i.e., the mode's (effective) amplitude, phase, frequency, and damping time. The values of $\omega_{\ell m}$ and $\tau_{\ell m}$ only depend on the mass and spin of the remnant black hole, but in this work we consider all frequencies as independent parameters in order to assess QNM detectability with more confidence, in the spirit of ``agnostic'' BH spectroscopy~\cite{Baibhav:2023clw}.

Inverting the FIM yields the covariance matrix
\begin{equation}
    \Sigma_{ij}=(\Gamma^{-1})_{ij}\ ,
\end{equation}
with diagonal (off-diagonal) components corresponding to the errors on (or correlations coefficients between) the parameters, i.e.,
\begin{equation}
\sigma_{\theta_i}=\sqrt{\Sigma_{ii}}\quad\ ,\quad 
c_{ij}=\Sigma_{ii}/(\sigma_{\theta_i} \sigma_{\theta_j})\ .
\end{equation}
To avoid numerical errors that can arise in attempting to invert a Fisher matrix with a very large condition number,
we use the method described 
in~\cite{Yagi:2009zm}, 
normalizing $\Gamma_{ij}$ to its 
diagonal components
before inverting it.

Due to the orthogonality properties of the spin-weighted spherical harmonics, the FIM for modes with different harmonic indices becomes block-diagonal. If $n=1,\ldots,k$ labels the QNMs included in the template, we have
\begin{equation}
    \Gamma_{ij}=\begin{pmatrix}
        \Gamma_{ij}^{(1)} & & & \\
         & \Gamma_{ij}^{(2)} & & \\
         & & \ddots & \\
         & & & \Gamma_{ij}^{(k)}
    \end{pmatrix}\ ,
\end{equation}
where $\Gamma_{ij}^{(1,2... k)}$ are the Fisher matrices of each individual QNM~\citep{Maselli:2023khq}.  

Modes with the same angular dependence require the calculation of cross terms which contain mixed derivatives. In our analysis, this only occurs for the $(44)$ and the $(22\times 22)$ quadratic mode. To determine the errors on the parameters of these modes we need to compute the $8\times8$ FIM for $\Vec{\theta}=(\bar{\mathcal{A}}_{44},\phi_{44},\omega_{44},\tau_{44},\bar{\mathcal{A}}_{22\times 22},\phi_{22\times 22},\omega_{22\times 22},\tau_{22\times 22})$.  The errors on these parameters are unaffected by the FIM elements related to modes with $(\ell m)\neq (44)$.

\subsection{Quasinormal mode starting time estimates}
\label{subsec:starting_time}

In order to evaluate the SNR and the statistical errors on the parameters of the quadratic QNMs, we first need to estimate the corresponding mode amplitude ${\cal A}_{\ell m}$, which appears in the waveform model of Eqs.~\eqref{eq:tildehp}-\eqref{eq:tildeh}.  The procedure requires a careful choice of the starting time $t_0$ after the peak of the signal. Choosing the value of $t_0$ effectively determines the onset of the perturbative regime, and is key to guaranteeing the validity of the QNM expansion.

There is no unique, unambiguous way of choosing $t_0$. Here we follow the approach introduced in Ref.~\cite{Baibhav:2017jhs}, which makes use of Nollert's ``energy maximized orthogonal projection'' (EMOP) criterion~\citep{nollert}. In this framework, the time-domain ringdown waveform $h$ is split into components ``parallel'' and ``perpendicular'' to the QNM. The energy ``parallel'' to a QNM is given by
\begin{equation}
    E_{||}=\frac{\omega_{\rm i}|\int_{t_0}\dot{h}_{\rm NR}\dot{h}^*|^2}{4\pi(\omega_{\rm r}^2+\omega_{\rm i}^2)}\ ,
    \label{eqn:EMOP}
\end{equation}
where we omit the angular indices $(\ell m)$ to simplify the notation, $\omega_{\rm r}$ and $\omega_{\rm i}$ denote the real and imaginary parts of the QNM frequency, respectively, and $h_{\rm NR}$ is a multipolar component of the strain extracted from numerical relativity simulations.
The ringdown starting time $t_0$ is then defined as the integration limit for which $E_{||}$ is maximized.
The integral above then provides the ``EMOP energy'' carried by the fundamental mode for a given multipole, $E_{\ell m}$. Reference~\cite{Baibhav:2017jhs} analyzed numerical relativity catalogs to find semianalytical fits of $E_{\ell m}$ as a function of the parameters of binary progenitors with aligned spins.

The energy $E_{\ell m}$ can then be mapped to the QNM amplitude by introducing the radiation efficiency
\begin{equation}
\epsilon_{\rm{RD}}=\frac{1}{M}\int_{f_{\rm{min}}}^{f_{\rm{max}}}\frac{dE}{df}df\ ,
    \label{eqn:rad_eff}
\end{equation}
where $dE/df$ is the gravitational wave energy spectrum, which is also related to the SNR $\rho$ by
\begin{equation}
\rho^2=\frac{2}{5\pi^2r^2}\int_{f_{\rm{min}}}^{f_{\rm{max}}}\frac{1}{f^2S_n(f)}\frac{dE}{df}df\ .
\label{eqn:SNR_dEdf}
\end{equation}
By comparing Eqs.~\eqref{eqn:SNR_dEdf} and~\eqref{eqn:snr}, we can express $dE/df$ as a function of the waveform parameters $({\cal A}_{\ell m},M_f,r,\phi_{\ell m},\omega_{\ell m},\tau_{\ell m})$, and solve for the EMOP-based amplitudes (hereafter ${\cal A}_{\ell m,{\rm E}}$) for a given choice of $\epsilon_{\rm{RD}}$, as determined by the fits of Ref.~\cite{Baibhav:2017jhs}.

This method allows us to estimate the amplitudes ${\cal A}_{\ell m,{\rm E}}$ just for a few linear modes. However, the authors of Ref.~\cite{Cheung:2023vki} recently derived fitting formulas for the amplitudes and phases of linear modes (including overtones and retrograde modes) as well as nonlinear modes. These fits, which are based on a fitting algorithm applied to over 500 binary BH simulations in the SXS catalog, give the amplitudes and phases as functions of the progenitor BH parameters: the binary mass ratio and the projections of the individual spins along the orbital angular momentum. We use these amplitude and phase fits for our SNR and error calculations, and we estimate the starting time $t_0$ by looking for the time at which the amplitude fits of Ref.~\cite{Cheung:2023vki} agree well with the EMOP estimate ${\cal A}_{\ell m,{\rm E}}$.

Our procedure is as follows. For each linear mode, we compute the percentage difference between the amplitude fit $\mathcal{A}_{\ell mn}$ of Ref.~\cite{Cheung:2023vki} and the EMOP fit ${\cal A}_{\ell m,{\rm E}}$ of Ref.~\cite{Baibhav:2017jhs} over a four-dimensional space including the three progenitor binary parameters -- mass ratio $q\equiv m_1/m_2$ in the range $[1,\,10]$, plus symmetric and antisymmetric spin combinations $\chi_+\equiv (q\chi_1+\chi_2)/(1+q)$ and $\chi_-\equiv (q\chi_1-\chi_2)/(1+q)$ in the range $[-0.99,\,0.99]$ -- and values of $t_0$ in the range $[0,\,30]\,M_f$. For each mode, we then select the points for which the combinations of these four parameters ($q, \chi_+, \chi_-, t_0$) result in a percentage difference between fits smaller than $5\%$. The values of $t_0$ that lead to such small percentage differences for the $(22)$ and $(33)$ fundamental modes are plotted in Fig.~\ref{fig:t_corr} as functions of $\chi_+$, for various selected values of $q$ and $\chi_-$. The plot shows that the values of $t_0$ at which the amplitude fits agree are roughly in the range $[10,\,15]\,M_f$, except at large values of $\chi_+$. The dark blue-green markers, corresponding to $q=1$, are absent in the bottom panel because $\mathcal{A}_{33}\rightarrow0$ in the equal-mass limit. While this is not shown in the figure for brevity, analogous plots for the linear $(21)$ and $(44)$ modes show that the optimal values of $t_0$ are similarly centered somewhere in the range $[10,\,15]\,M_f$.

\begin{figure}[t]
 \includegraphics[width=0.45\textwidth]{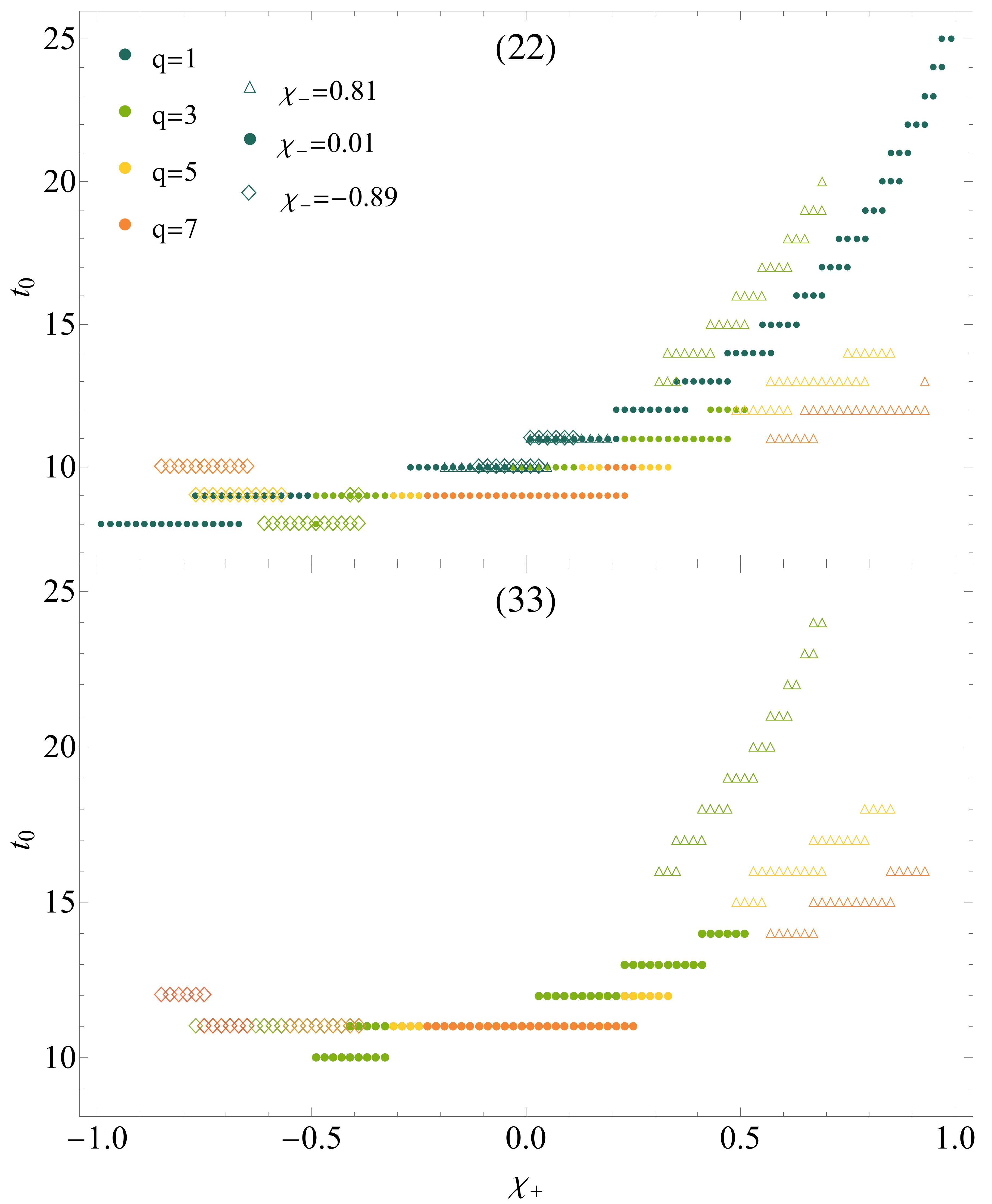} 
 \caption{Results of our estimates of the optimal starting time $t_0$, based on comparisons between $\mathcal{A}_{\ell m,E}$ and $\mathcal{A}_{\ell mn}$. For each mode, we fix $q$ and $\chi_-$, and we plot the values of $t_0$ for which the difference between the two amplitude fits is smaller than $5\%$ as a function of $\chi_+$. Different colors correspond to different values of $q$. Different point marker shapes (triangles, dots, and diamonds) correspond to selected values of $\chi_-=0.81$, $0.01$, and $-0.89$, respectively. }
 \label{fig:t_corr}
\end{figure}

Armed with this qualitative understanding, we then estimate $t_0$ for each mode by averaging over all the times which provide an agreement between amplitude fits better than $5\%$. These averaged values of $t_0$ are summarized in Table~\ref{tbl:t_corr}. Overall, these results suggest that, when accounting for the amplitude dependence on the mass ratio and on the spin values, the agreement between fits is maximized around $t_0\approx12M_f$. This is the value we will use hereafter for SNR and Fisher calculations.

\begin{table}[b]
\caption{\label{tbl:t_corr} 
  Optimal ``averaged'' starting time $t_0$ for QNMs with different $(\ell m)$, evaluated through the procedure outlined in Sec.~\ref{subsec:starting_time}.
}
\begin{ruledtabular}
\begin{tabular}{lcdr}
{}& Mode & \multicolumn{1}{c}{``Optimal'' $t_0$ }& \\
\colrule
    &(22) & 10.62 & \\
    &(21) & 12.21 & \\
    &(33) & 12.66 & \\
    &(44) & 10.94 & \\ 
\end{tabular}
\end{ruledtabular}
\end{table}

\begin{figure}[t]
 \includegraphics[width=0.48\textwidth]{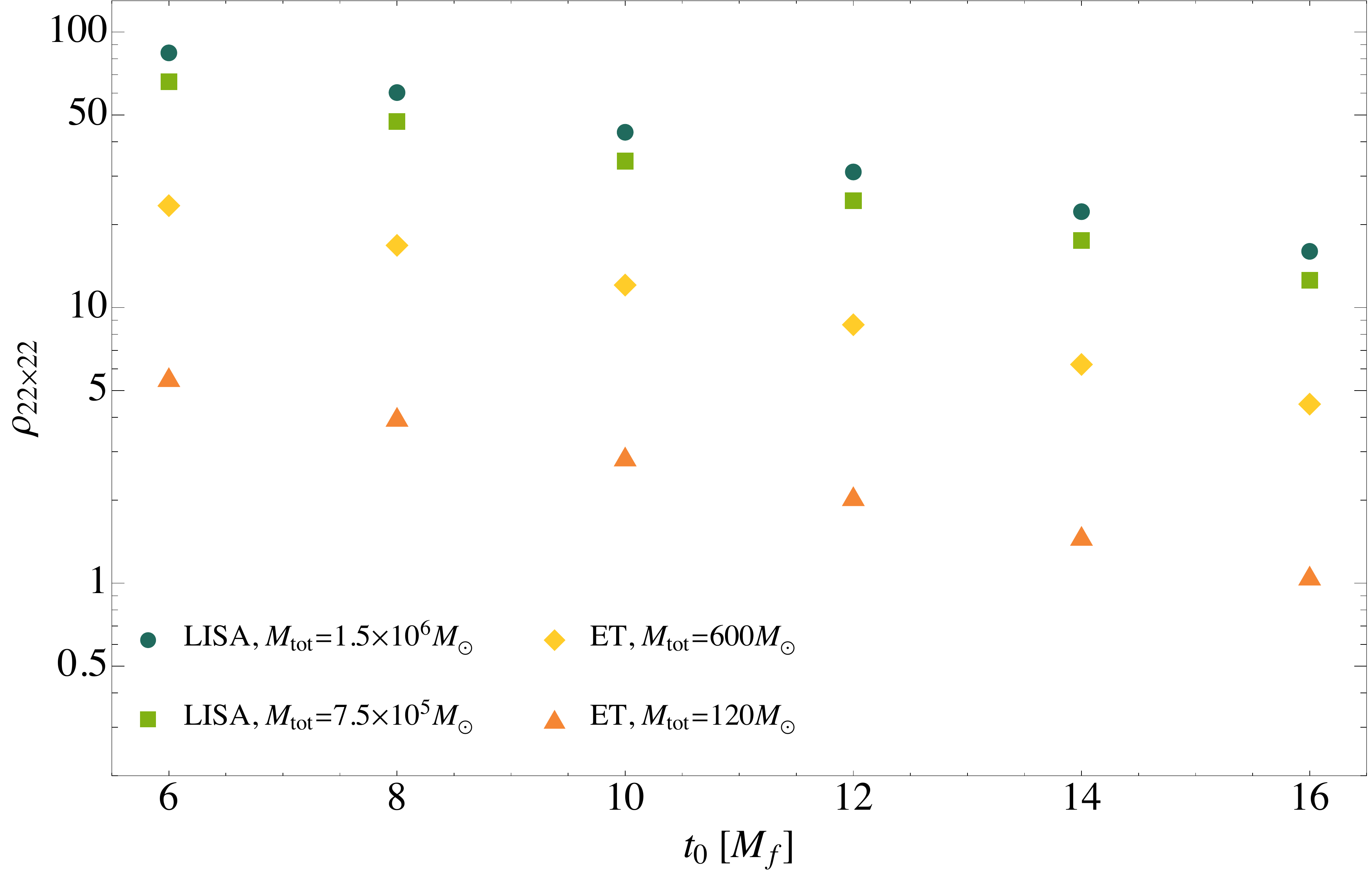} 
 \caption{SNR of the $(22\times 22)$ quadratic mode as a function of the starting time $t_0$. Different bullet points refer to BHs with different masses, observed by LISA or ET. All sources assume nonspinning progenitors, with a mass ratio $q=2$. Events detected by LISA and ET are located at  $z=10$ and $z=1$, respectively. 
}
 \label{fig:amp_errors_diff_t0}
\end{figure}

While this choice is somewhat arbitrary, we can understand if the SNR and the FIM-estimated statistical errors are sensitive to the choice of starting time by varying $t_0$ around our fiducial value. In Fig.~\ref{fig:amp_errors_diff_t0} we show the SNR of the quadratic mode $\rho_{22\times22}$ as a function of $t_0$ for a few selected binary progenitor parameters (see Appendix~\ref{app:bracket} for a more comprehensive analysis).

While the starting time can certainly affect the estimates, from Fig.~\ref{fig:amp_errors_diff_t0} we see that variations in $t_0/M_f$ of order unity result in variations in SNR (and hence, variations in parameter estimation errors) of order unity. This can be understood through the following simple considerations. To compute the SNR at a given $t_0$, we multiply each mode in the time-domain waveform \eqref{eq:h_ringdown} by $e^{i(\omega_{\ell m} + i/\tau_{\ell m})t_0}$.  Absorbing the real and imaginary parts of this time correction into the QNM phase and amplitude, respectively, the time-shifted QNM parameters become
\begin{eqnarray}\label{eqn:t0}
    \Bar{\mathcal{A}}_{\ell m} \longrightarrow \Bar{\mathcal{A}}_{\ell m} e^{-t_0/\tau_{\ell m}}\quad \ ,
    \quad\phi_{\ell m} \longrightarrow \phi_{\ell m} + \omega_{\ell m} t_0\ .
\end{eqnarray} 
All calculations in the frequency domain can then proceed along the steps described in Sec.~\ref{sec:templates}.
Typical values of 
$t_0/\tau_{\ell m}$ are less than 2, 
such that the exponential factor 
$e^{-t_0/\tau_{\ell m}}$ does not affect the 
SNR (and the FIM errors) dramatically. 

Let us stress once again that the criterion for determining the starting time of the quadratic mode laid out in this section is only a rough estimate.  A better motivated criterion could be to use the earliest time at which the bias in the frequency, amplitude and phase of the quadratic mode is lower than the statistical error of measurement.  Such a criterion is discussed, e.g., in Refs.~\cite{Cheung:2023vki,Clarke:2024lwi}.  In principle, this can be done on a per-event basis, as follows: (i) First pick an initial guess of $t_0$, and use a FIM calculation to estimate the statistical error;  (ii) To estimate the systematic bias (due to the fact that we do not start at sufficiently late times), pick a numerical relativity simulation with parameters close to the event, fit the waveform at $t_0$, and compare the result with an estimate of the (unbiased) value obtained from fits performed at a time at which the frequency, amplitude and phases are the most stable (see Ref.~\cite{Cheung:2023vki}); (iii) Iteratively tune $t_0$ until the statistical error and the systematic bias are of similar magnitude. The resulting $t_0$ would be the optimal starting time to use for the event in question.  Here we do not apply this procedure because it is prohibitively expensive, especially for the large astrophysical catalogs considered below. Given the order-of-magnitude nature of FIM calculations and the uncertainties in the astrophysical event rates of binary BH mergers, the simple EMOP estimates of $t_0$ used here are sufficiently accurate.

\section{Astrophysical catalogs}
\label{sec:catalogs}

To estimate the rates of binary BH merger events with detectable nonlinear modes, in this section we introduce astrophysical models of binary BH populations that fall within the observational window of XG detectors.

In Sec.~\ref{sec:SOBH} we focus on ground-based detectors, such as CE and ET. In this case we consider populations compatible with the third LIGO-Virgo-KAGRA observing run with two different assumptions on the component spins (\texttt{model~I} and \texttt{model~II}, also used in Ref.~\cite{Maselli:2023khq}), as well as a population of dynamically formed BH binaries (\texttt{model~III}).
In Sec.~\ref{sec:MBH_description} we introduce a range of plausible massive black hole (MBH) population models that can produce ringdown events detectable by LISA.

\subsection{Stellar mass binary black hole population models}
\label{sec:SOBH}

We consider two astrophysical populations with masses sampled from the POWER LAW$+$PEAK model motivated by the third LIGO-Virgo-KAGRA gravitational wave transient catalog, GWTC-3~\cite{KAGRA:2021duu}. The binaries are drawn up to redshift $z=10$ from a distribution which follows the Madau-Dickinson cosmic star formation rate (SFR), with a local binary BH merger rate ${\cal R}_{\rm m}= 28.3 {\rm Gpc}^{-3}{\rm yr}^{-1}$~\cite{Madau:2014bja,Ng:2020qpk}.  Following Ref.~\cite{Maselli:2023khq}, we consider two possible prescriptions for the spins of the component BHs, which are always assumed to be either aligned or antialigned with the orbital angular momentum of the binary. In \texttt{model~I} the dimensionless spin parameters are sampled from a Beta distribution with ($\alpha=2,\, \beta=5$) peaked around 0.2. In \texttt{model~II} the BH spins are sampled from a uniform distribution within $[-1,1]$ (see Sec.~IIE of~\cite{Maselli:2023khq} for further details).

We also consider a third model (hereafter \texttt{model~III}) consisting of merger events simulated using {\tt Rapster}, a population synthesis code for binary BH mergers produced dynamically in globular clusters~\cite{Kritos:2022ggc}.
The star cluster formation rate follows the redshift distribution in Eq.~(11) of Ref.~\cite{Mapelli:2021gyv} in the redshift range $z\in [0,\,10]$.
The BH masses are generated from the {\tt SEVN} code~\cite{Spera:2017fyx}, and the stellar masses of BH progenitors follow the Kroupa initial mass function.
Since the BH mass spectrum depends strongly on the metallicity of the system, given the redshift of cluster formation, we sample the metallicity from a lognormal distribution with a mean value given by Eq.~(6) of Ref.~\cite{Madau:2016jbv} and standard deviation of 0.25; these assumptions are motivated by Ref.~\cite{Mapelli:2021gyv}.
Current gravitational wave observations constrain the BH spins to be relatively low~\cite{KAGRA:2021duu}. We assume, somewhat arbitrarily, that the spin magnitudes of first-generation BHs (those that form from the collapse of massive stars) are uniform in the range $[0, 0.2]$.
Based on observations of young clusters in the local Universe~\cite{2019ARA&A..57..227K}, we draw the mass of star clusters $M_{\rm cl,0}$ from a Schechter initial mass function with spectral index $-2$ and truncation mass scale at $10^7M_\odot$ in the range $[10^4,10^8]M_\odot$.
The initial half-mass radius $r_{\rm h,0}$ (a required input parameter in the {\tt Rapster} code) is weakly correlated with $M_{\rm cl,0}$, and it is sampled from a lognormal distribution with mean given by Eq.~(3) of~\cite{Larsen:2003fp} and a scatter of $0.7\,\rm dex$.
Finally, the initial galactocentric radius of clusters is drawn from a S\'ersic profile with index $n=1$ and scale radius $1\,\rm kpc$ (a typical Milky Way-like galaxy).
All other {\tt Rapster}  input parameters are set to their default values (see Table~1 of~\cite{Kritos:2022ggc}). In Appendix~\ref{app:merger_rate} we discuss the merger rates estimated with {\tt Rapster}, along with the sampling technique we use to draw events from the binary BH population generated by the code.

\subsection{Massive black hole evolution models}
\label{sec:MBH_description}

The evolution of MBHs and of their host galaxies is followed using the semianalytic model introduced in Ref.~\cite{Barausse:2012fy} and refined in later studies~\cite{Sesana:2014bea,Antonini:2015cqa,Antonini:2015sza,Bonetti:2017dan,Bonetti:2018tpf,Barausse:2020mdt}. Galaxies are comprised of dark matter halos accreting gas from the intergalactic medium. This chemically primordial gas can either flow directly to the core of the halo along cold filaments~\cite{Dekel2006,Cattaneo:2006rp,Dekel:2008em} (either at high redshift, or in small halos at low redshift), or it can get shock-heated to the halo's virial temperature, before cooling and settling at the center. By conservation of angular momentum~\cite{1998MNRAS.295..319M}, this cold gas can form disk structures~\cite{1998MNRAS.295..319M}, becoming a site for star formation.  When these disks get disrupted by bar instabilities of major galaxy mergers, star formation bursts occur and spheroidal structures (bulges) form.  The semianalytic model also incorporates smaller structures such as nuclear star clusters~\cite{Antonini:2015cqa,Antonini:2015sza}, a central gas reservoir for MBH accretion~\cite{Granato:2003ch}, besides the MBHs themselves, for which the mass and spin evolution is followed through their accretion and merger history. The models also account for feedback processes on the growth of structures, namely the impact of active galactic nuclei (AGNs), which expel/heat up gas from the galactic centers in high mass systems~\cite{Croton:2005hbr,2008ApJS..175..390H,2006MNRAS.370..645B}, potentially quenching star formation and MBH accretion, and the damping effect of supernova (SN) explosions on star formation in shallow potential wells~\cite{Springel:2001qb,Fujita:2004mp,Rasera:2005gq}. SN feedback can also suppress MBH accretion and the gas-driven hardening of MBH binaries in small systems with escape velocities $\lesssim 270$ km/s~\cite{vanMeter:2010md}, lower than the typical speed of SN winds~\cite{2017MNRAS.468.3935H}.

The dark-matter merger tree on which the model relies to describe the halo merger history is produced via an extended Press-Schechter formalism~\cite{Press:1973iz}, calibrated against results from $N$-body simulations~\cite{Parkinson:2007yh}. Galaxy mergers follow halo mergers, but with potentially large delays. First, one has to account for the initial survival of the smaller halo within the larger one as a sub-halo (or satellite), which slowly sinks in by dynamical friction, while being tidally disrupted and evaporating~\cite{Boylan-Kolchin:2007bvo,Taffoni:2003sr}. Besides this first delay, the baryonic components (the galaxy proper) are also subject to dynamical friction and tidal disruption/evaporation~\cite{2008gady.book.....B}. MBH mergers experience additional delays from the galaxy/halo merger, as MBH pairs need to travel all the way from separations of hundreds of pc or even kpc (where they are expected to be at the coalescence of their host galaxies) down to $\sim$ pc scales, where they form bound binaries, and eventually to the sub-pc scales on which gravitational wave emission alone can drive them to merger.  At separations of hundreds of pc, the main driver of the evolution is again dynamical friction~\cite{2017ApJ...840...31D}, calibrated to the results of hydrodynamic simulations~\cite{2018MNRAS.475.4967T}, while the evolution from pc to sub-pc scales is driven by stellar hardening~\cite{Quinlan1996,Sesana:2015haa}, interactions with gas (if available)~\cite{Macfadyen:2006jx,Cuadra:2008xn,Lodato:2009qd,Rodig:2011jz,Nixon:2010by,Duffell:2019uuk,Munoz:2018tnj} and/or with other MBHs from previous galaxy mergers~\cite{Hoffman:2006iq,Bonetti:2016eif,Bonetti:2017dan,Bonetti:2017lnj,Bonetti:2018tpf}.  Upon merger, the model updates the MBH mass and spin based on semianalytic formulas reproducing the results of numerical relativity simulations~\cite{Barausse:2012qz,Hofmann:2016yih} and includes the impact of gravitational wave recoil~\cite{vanMeter:2010md}, which can lead to the ejection of the merger remnant from the galaxy.

In addition to the delays between halo/galaxy mergers and MBH mergers, the model's predictions for the MBH binary population are deeply impacted by the unknown initial BH seed mass function at high redshift. In the light seed (LS) scenario, MBHs are assumed to originate from seeds of a few hundred $M_\odot$, remnants of Pop~III stars exploding as SNe. Specifically, the model populates large halos forming from the $3.5\sigma$ peaks of the primordial density field at $z\gtrsim 15$ with seed BHs, estimating their mass to be $\sim 2/3$ of the initial Pop~III mass (to account for mass loss during the SN explosion). The star's initial mass distribution is modeled as a lognormal distribution centered at $300 M_\odot$ with a standard deviation of 0.2 dex, excluding masses between $140$ and $260\,M_\odot$, which produce pair-instability SNe leaving no BH remnant.  Our model also considers a heavy seed (HS) scenario, where MBHs form with mass already of $\sim 10^5 M_\odot$. This scenario is based on the model of Ref.~\cite{Volonteri:2007ax}, where seeds emerge from the collapse of protogalactic disks, driven by bar-instability in high-redshift, low-spin, and low-temperature halos. The seed masses are determined according to specific formulas from the same reference. It is of course conceivable that a combination of both HSs and LSs forms in nature, a hypothesis that LISA will help clarify~\cite{Toubiana:2021iuw}, but in this paper we will consider one scenario at a time.

The semianalytic model is calibrated against a number of galactic/subgalactic scale observables (e.g. galaxy, MBH baryonic mass functions, star formation rates and densities, scaling relations for MBHs and nuclear star clusters, AGN luminosity function, etc.)~\cite{Barausse:2012fy,Sesana:2014bea,Antonini:2015cqa,Antonini:2015sza,Bonetti:2017dan,Bonetti:2018tpf,Barausse:2020mdt} and also against the recent observations of a putative stochastic background of gravitational waves from MBH binaries by pulsar timing array experiments~\cite{Antoniadis:2023ott,Tarafdar:2022toa,NANOGrav:2023gor,Reardon:2023gzh,Xu:2023wog}. In more detail, we will consider here a subset of the models presented in Refs.~\cite{Barausse:2023yrx,EPTA:2023xxk}, where a comparison to pulsar timing array data is presented.  Despite this calibration, the uncertainty on the predictions of the merger rate of MBH binaries observable by LISA remains significant.

In this paper, we bracket these uncertainties by focusing on six models from Refs.~\cite{Barausse:2023yrx,EPTA:2023xxk} predicting rather different LISA event rates. In particular, we will consider models ``HS-nod-noSN (B+20),'' ``LS-nod-noSN (B+20),'' ``LS-nod-SN (B+20),'' ``popIII-d (K+16),'' ``Q3-d (K+16),'' and ``Q3-nod (K+16),'' where ``HS/Q3'' (``LS/popIII'') indicate models with heavy (light) seeds, respectively; ``nod'' indicates models with no delays between galaxy and MBH mergers (except for the dynamical friction time -- including tidal effects -- between halos); ``d'' indicates models which include not only the delays between halo and galaxy mergers, but also for stellar hardening, MBH triplets and gas-driven migration; and ``(K+16)/(B+20)'' refer to the papers where the models were first presented (Refs.~\cite{Klein:2015hvg,Barausse:2020mdt}). 
``SN'' and ``noSN'' refers to whether the quenching effect of SN winds 
on MBH accretion is accounted for or not, respectively.
Note that Refs.~\cite{Barausse:2023yrx,EPTA:2023xxk} presented 
predictions of each of these models at finite merger tree resolution, and extrapolated predictions to infinite resolution. As discussed in Refs.~\cite{Barausse:2023yrx,EPTA:2023xxk}, the extrapolated results should be considered as upper limits to the predicted merger rates, while the finite-resolution results provide lower limits.

\section{Detectability of nonlinear modes}
\label{sec:detectability}

To understand which events in our catalogs are likely to have a detectable quadratic mode, we begin by computing the SNR for this mode alone. We will then select the events with detectable nonlinear modes (i.e., those with $\rho_{22\times22}>8$) for further analysis. 

\begin{figure}[t]
\centering
\includegraphics[width=0.49\textwidth]{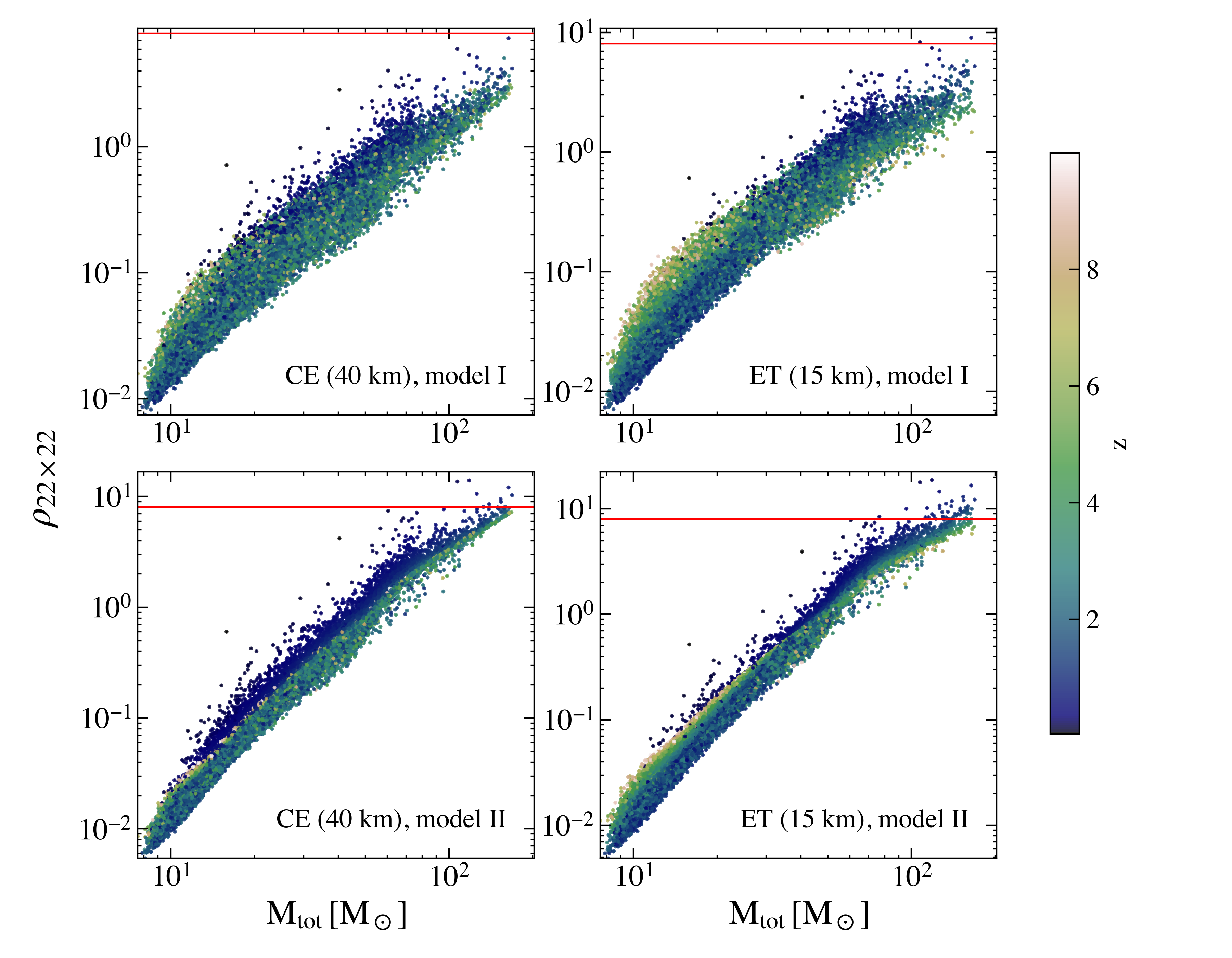} 
\caption{SNR of the (22$\times$22) quadratic mode computed for \texttt{model~I} and \texttt{model~II} (POWER LAW$+$PEAK mass distribution and two different spin distributions). Plot points are colored by redshift, $z$.
The horizontal red line in each plot marks our detectability threshold of $\rho_{22\times22}=8$.}
\label{fig:parspec}
\end{figure}

\begin{figure}[t]
\centering
\includegraphics[width=0.48\textwidth]{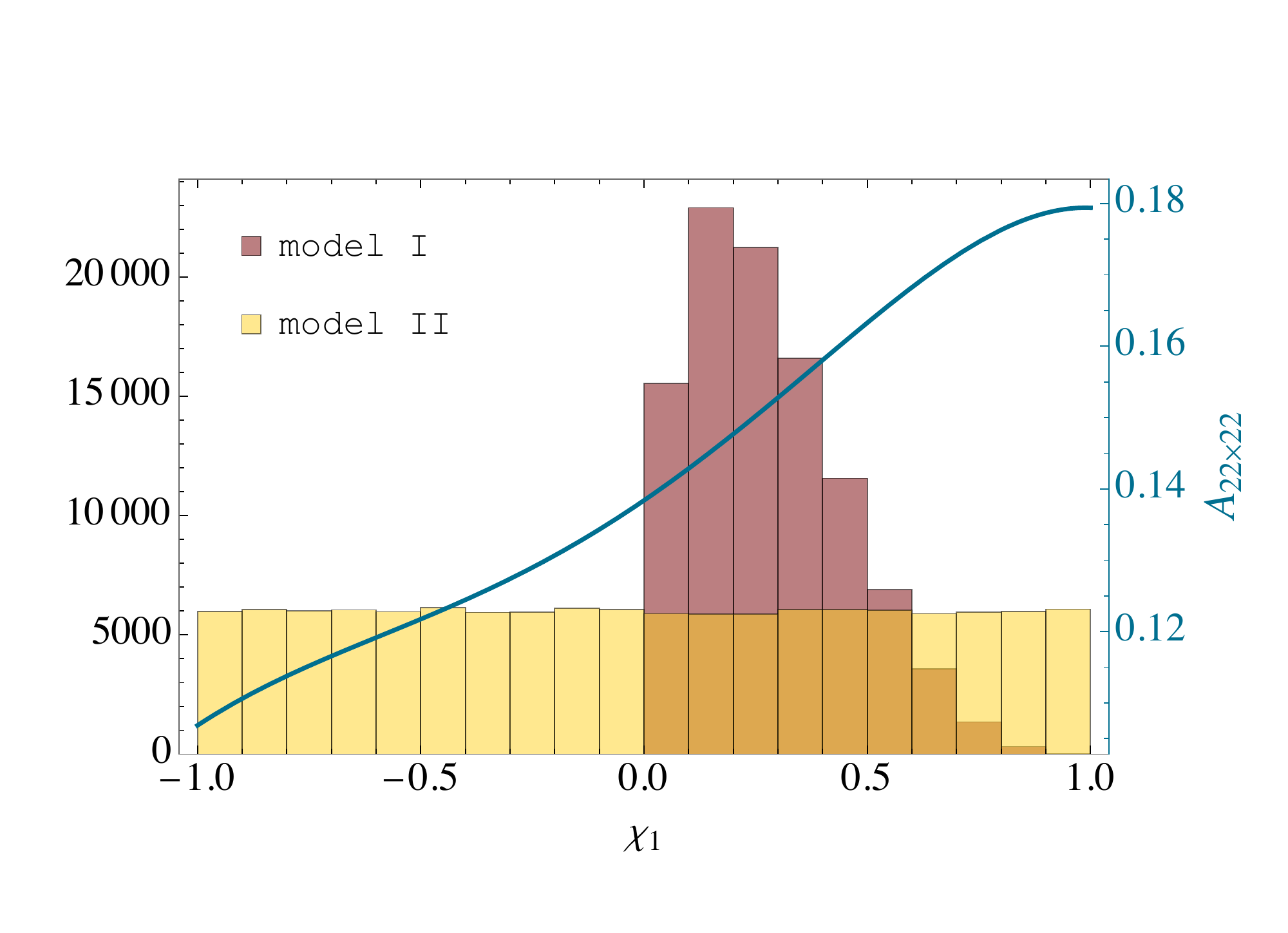} 
\caption{Quadratic amplitude dependence on spin. This plot explains why \texttt{model~II} has a larger number of events with detectable quadratic modes, as compared to \texttt{model~I}. Histograms (left axis): number of events in the \texttt{model~I} and \texttt{model~II} catalogs with a given value of $\chi_1$ (which in the catalogs is assumed to be equal to $\chi_2$). Teal line (right axis): quadratic mode amplitude for equal-mass mergers ($q=1$) as a function of $\chi_1$. The qualitative behavior for unequal-mass binaries ($q>1$) is similar.}
\label{fig:spin_mod}
\end{figure}

\subsection{Quadratic mode SNR of stellar-mass black hole binaries as observed by ground-based detectors}\label{sec:parspec}
 
We first compute the quadratic mode SNR of $10^5$ binaries at $t_0=12M_f$ after the peak of the waveform for both CE and ET. 
We use the same catalogs (\texttt{model~I} and \texttt{model~II}) studied in Ref.~\cite{Maselli:2023khq}.

The results are shown in Fig.~\ref{fig:parspec}, where the color of the points corresponds to the redshift of each binary.
For each of the two detectors, the vast majority of the events have a quadratic mode SNR less than 8, with \texttt{model~II} yielding slightly larger SNRs than \texttt{model~I}.
This is because \texttt{model~II} has more events with large, positive progenitor spins, for which the quadratic mode amplitude is larger. This is illustrated in Fig.~\ref{fig:spin_mod}, where we plot the number of events with a given value of $\chi_1$ (left axis) and the magnitude of $\mathcal{A}_{22\times22}$ (right axis) as a function of the BH spin $\chi_1$ (note that each event in the \texttt{model~I} and \texttt{model~II} catalogs has $\chi_1=\chi_2$ by assumption). The quadratic mode amplitude is clearly higher as the individual spins approach the extremal limit. The quadratic mode amplitudes plotted in Fig.~\ref{fig:spin_mod} refer to equal-mass binaries ($q=1$), but we have checked that the behavior is qualitatively similar as long as $q\leq 3.6$, the largest mass ratio for binaries in the \texttt{model~I} and \texttt{model~II} catalogs.

Overall, under the conservative assumption that the mass distribution follows current LIGO-Virgo-KAGRA observations and that the redshift distribution follows the star formation rate, we conclude that the quadratic mode may be observable by XG detectors (in the sense that $\rho_{22\times 22} > 8$) for at most a few tens of stellar mass binary BH merger events per year. 

\begin{figure}[t]
\includegraphics[width=0.5\textwidth]{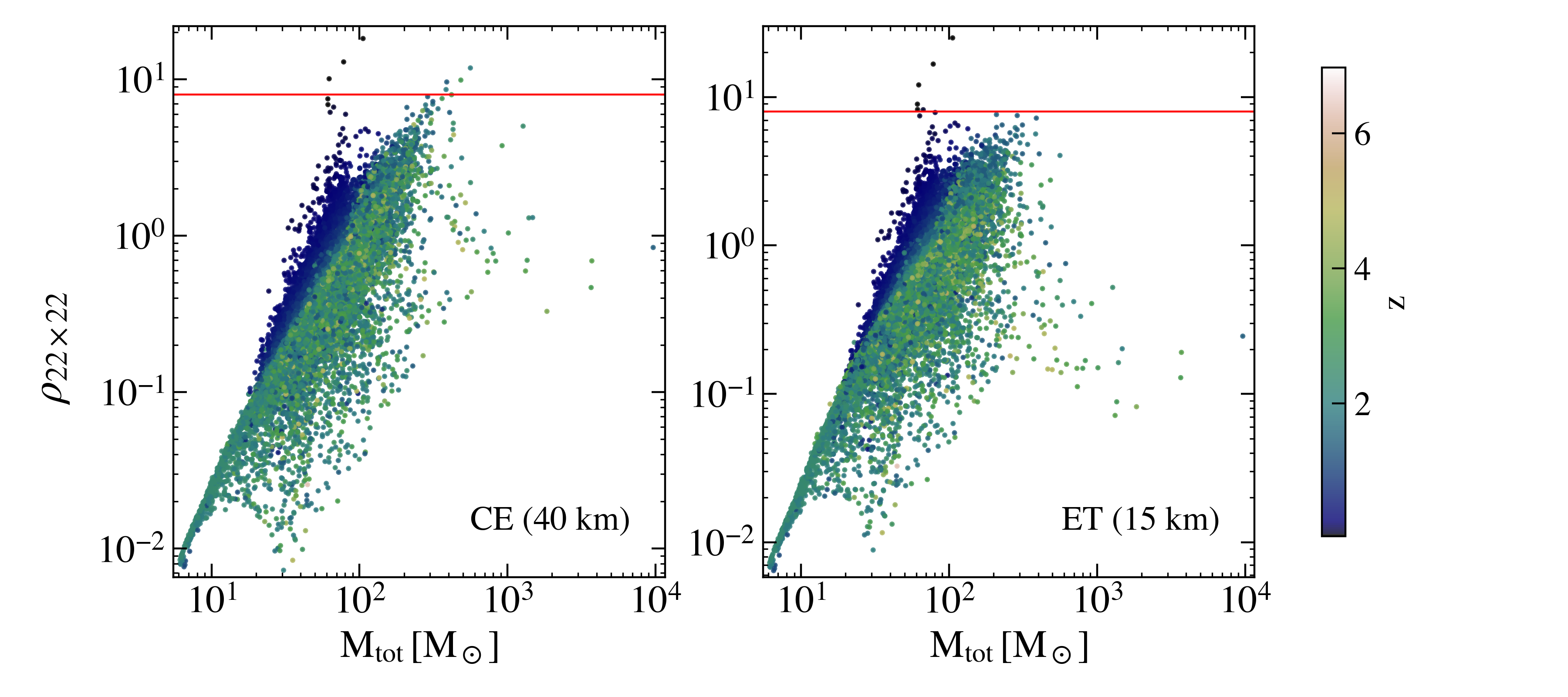} 
\caption{SNR of the quadratic mode  ($22\times22$) calculated for $10^5$ binary BH mergers simulated by \texttt{Rapster}, including numerous IMBHs, for both CE (left panel) and ET (right panel). The horizontal red line marks the detection threshold ($\rho_{22\times22}=8$).}
\label{fig:rapster_weighted}
\end{figure}

\begin{figure}[t]
\includegraphics[width=0.45\textwidth]{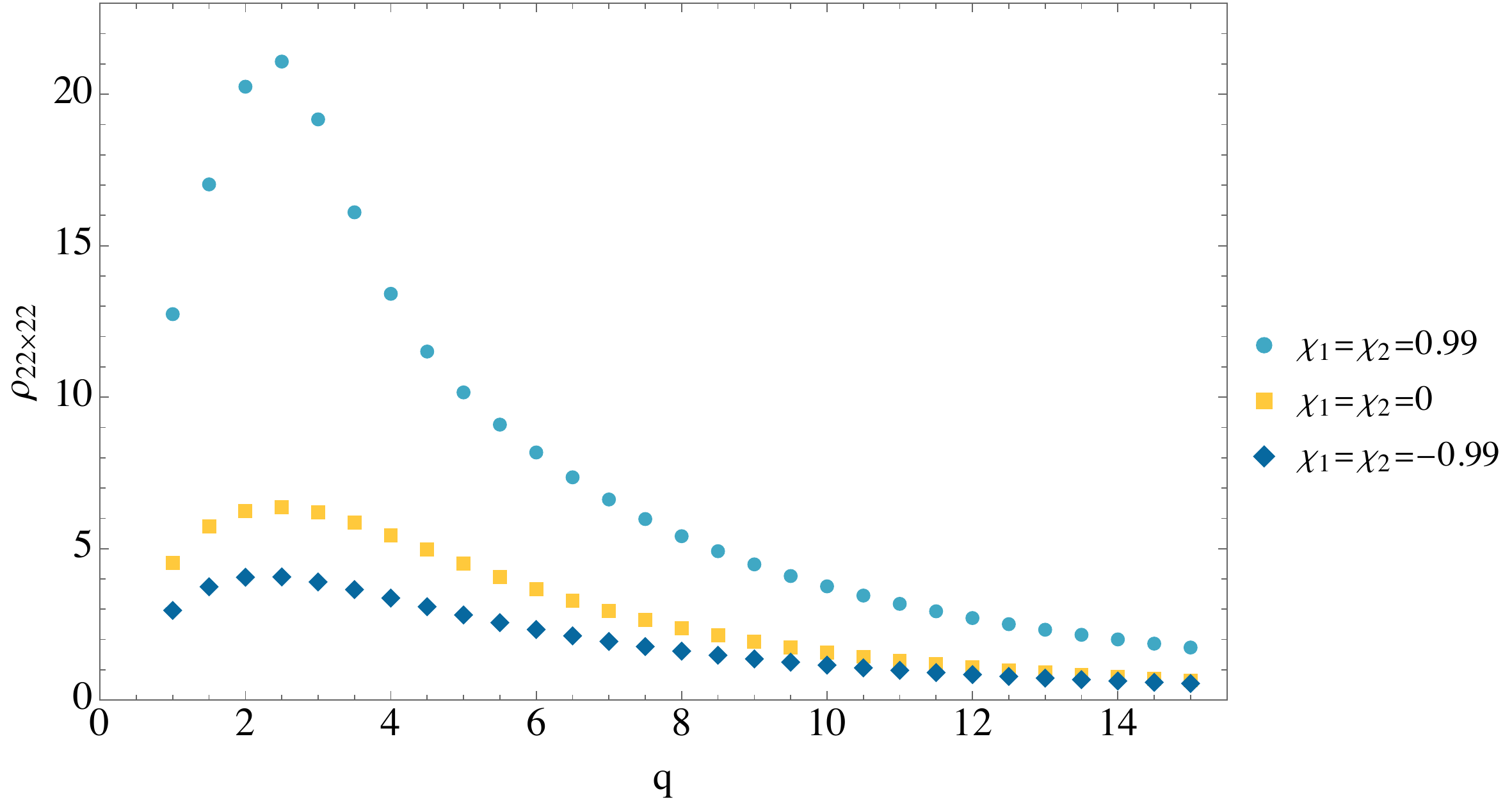} 
\caption{The quadratic mode SNR is lower at large values of $q$. Results are shown for CE's noise curve ($z=1$) and $m_2$ fixed to 100$M_\odot$, but results are qualitatively similar for other detectors and secondary mass values.}
\label{fig:snr_vs_q}
\end{figure}

\subsection{Quadratic mode SNR of dynamically formed black hole binaries as observed by ground-based detectors}\label{sec:rapster}

We now investigate whether prospects improve if we compute the quadratic mode SNR for binaries formed exclusively in dynamical channels. We simulate dynamical formation with the \texttt{Rapster} code~\cite{Kritos:2022ggc}, which allows for heavy binary mergers with primary masses $\mathcal{O}(100) M_\odot$.

The amplitude fits that we employ are not accurate for high mass ratios, so we conservatively exclude all events in the  \texttt{Rapster} catalog with $q>10$ (these constitute less than 0.7\% of all events). Since the quadratic mode amplitude, and the corresponding SNR, is significantly higher for comparable-mass binaries (see Fig.~\ref{fig:snr_vs_q}), our forecasts of quadratic mode detectability are only mildly affected by the implementation of this mass ratio cutoff. 

We take weighted samples of $10^5$ events at a time from the \texttt{Rapster} catalog of dynamically formed binary BH events, weighting samples by a realistic merger rate as a function of redshift (see Appendix~\ref{app:merger_rate}). The quadratic mode SNRs, again for a 40-km CE and a 15-km ET, are plotted in Fig.~\ref{fig:rapster_weighted}.  Once again, we find only a few events with a potentially detectable $\rho_{22\times22}$. For CE (ET), the mean quadratic mode SNR for the events shown in Fig.~\ref{fig:rapster_weighted} is 0.64 (0.81), with 7 (6) events having $\rho_{22\times22}> 8$. We have repeated the analysis several times, taking multiple weighted samples of $10^5$ events from the total catalog and computing the quadratic mode SNR for these events, and in each case we find qualitatively similar results.

Unlike the \texttt{model~III} binaries predicted by the \texttt{Rapster} simulations, which can have total mass $M_{\rm{tot}}\sim [300,\,600]\, M_\odot$, the most massive binaries in the \texttt{model~I} and \texttt{model~II} catalogs have a total mass of $\sim170\,M_\odot$.
In the particular \texttt{Rapster} universe realization shown in Fig.~\ref{fig:rapster_weighted}, all six of the events detectable by ET have masses $M_{\rm{tot}}\sim [60,\,100]\,M_\odot$ and redshifts $z\sim [0.04,\,0.07]$. For CE we find a total of seven detectable sources: three of them have values of $(M_{\rm tot},z)$ comparable to those observed by ET, while the remaining four have larger masses ($M_{\rm{tot}}\sim [300,\,600]\, M_\odot$) and are located at redshifts $z\sim [1.4,\,3]$. The improved sensitivity at lower frequencies allows CE to detect signals from heavier BHs.

\begin{figure}[t]
\includegraphics[width=0.48\textwidth]{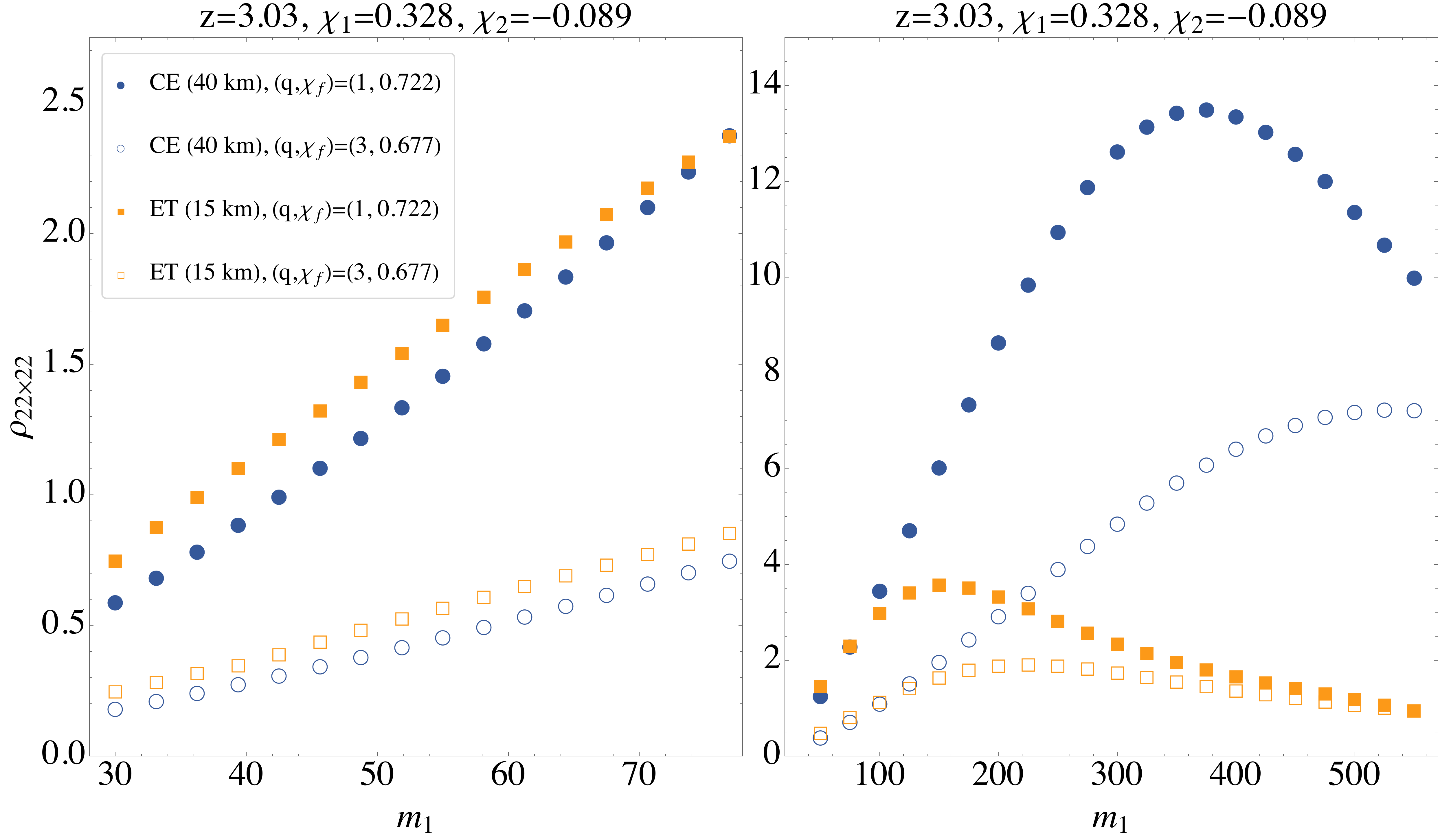} 
\caption{SNR of the $(22\times22)$ quadratic mode calculated for ET (orange) and CE (blue), as a function of the binary's primary mass, for mass ratio $q=1$ (filled markers) and $q=3$ (empty markers), and selected values of the component spins $\chi_{1,2}$ (these map into different values of the remnant spin, as indicated in the legend). The left and right panels refer to systems in the stellar-mass and IMBH range, respectively.
}
\label{fig:ET_vs_CE}
\end{figure}

This can be seen in Fig.~\ref{fig:ET_vs_CE}, where we plot the quadratic mode SNR as a function of the primary mass for binary BHs in the stellar-mass and IMBH ranges (left and right panels, respectively). While ET is slightly better at detecting the quadratic mode of binaries with individual masses $\mathcal{O}(10)M_\odot$, CE performs significantly better in the IMBH range.
These results suggest that binary BHs with masses somewhat larger than the range covered in the \texttt{model~I} and \texttt{model~II} catalogs may be good candidates for detection of the $(22\times22)$ quadratic mode with CE.

In Fig.~\ref{fig:hist_rapster_and_parspec} we show the distribution of detectable quadratic mode SNRs ($> 8$) for each of the catalogs discussed thus far over one year of observation. The astrophysical models used here suggest that we may expect at most $\mathcal{O}(10)$ events with a detectable $(22\times22)$ mode during each year of operation of ground-based XG detectors. 

\begin{figure}[t]
\includegraphics[width=0.44\textwidth]{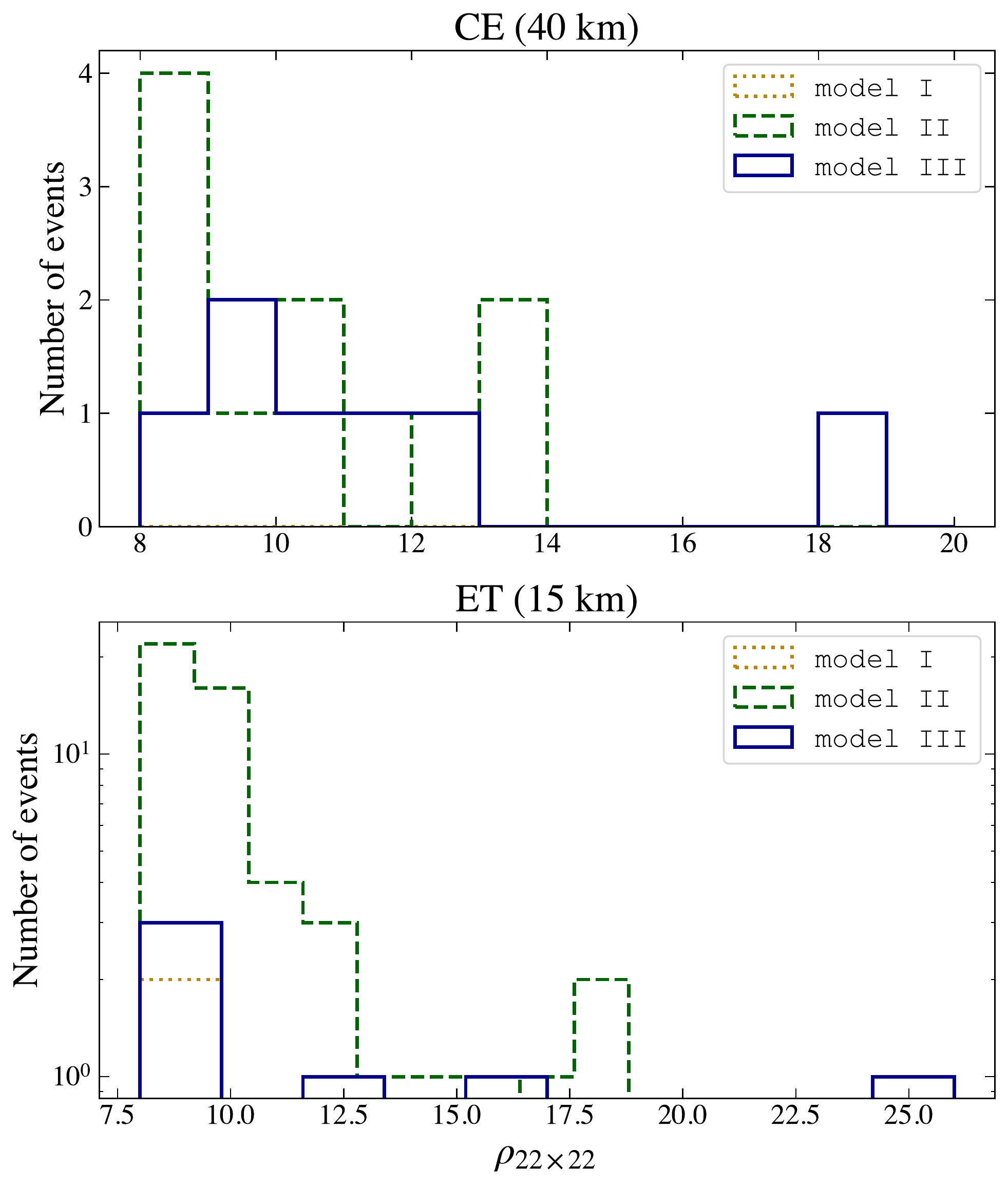} 
\caption{Quadratic mode SNR distribution of observable events ($\rho_{22\times22}>8$) over one year of operation of ground-based XG detectors. Note that in the top panel we use a linear scale, while in the bottom panel we use a log scale.}
\label{fig:hist_rapster_and_parspec}
\end{figure}

\subsection{Massive black hole observations with LISA}\label{sec:MBH_snrs}

\begin{figure}[t]
\includegraphics[width=0.45\textwidth]{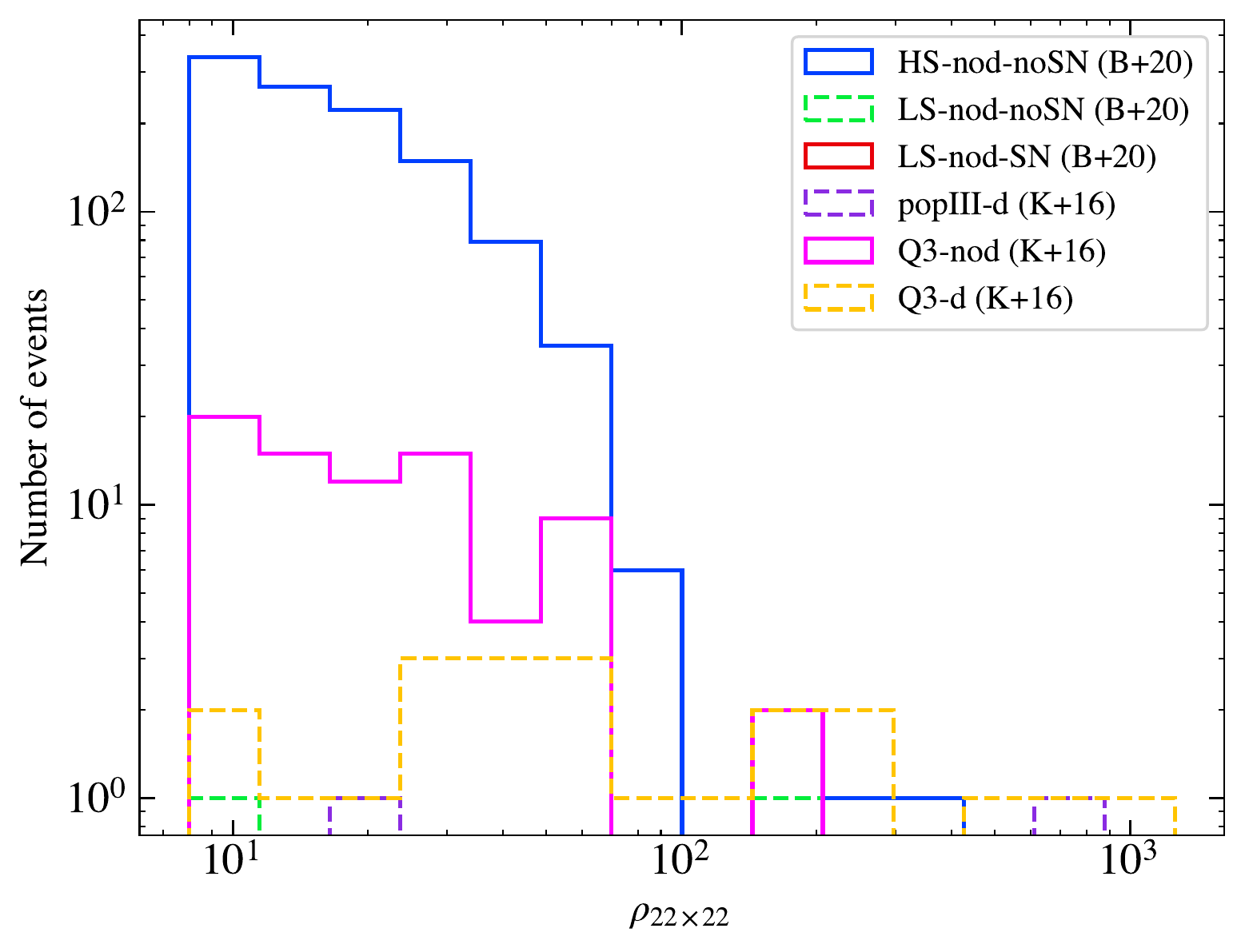}
\caption{SNR distribution of MBH events with a detectable quadratic mode ($\rho_{22\times22}>8$) over 4 years of LISA observations. Results are shown for the finite-resolution catalogs; extrapolated results show a similar distribution of events, but with rates increased by a factor of a few.}
\label{fig:MBH_hist}
\end{figure}

We now compute the quadratic mode SNRs for the six astrophysical models from Ref.~\cite{Barausse:2023yrx} discussed in Sec.~\ref{sec:MBH_description}. We use the LISA noise curve~\cite{Robson:2018ifk}, and again we take the starting time of the modes to be $t_0=12M_f$ after the peak of the waveform.

The results for each model are plotted in Fig.~\ref{fig:MBH_hist} for a single realization of a 4-year LISA observation time using models with finite merger tree resolution, as described in Sec.~\ref{sec:MBH_description}. In Table~\ref{tbl:MBH_stats_fin_res} we list the mean and maximum values of $\rho_{22\times 22}$ for each model, as well as the number of events with $\rho_{22\times 22}>8$ expected to be observed in four years. Alongside each value, we report in parentheses the corresponding value for models with extrapolated merger rates (the ``infinite resolution'' models described in Sec.~\ref{sec:MBH_description}). For context, for each model, we also display the number of events with fundamental mode SNR $\rho_{22}>8$, and with full inspiral-merger-ringdown (IMR) ${\rm SNR}>8$. We again exclude events with $q>10$ from both Fig.~\ref{fig:MBH_hist} and Table~\ref{tbl:MBH_stats_fin_res}.
Whereas Fig.~\ref{fig:MBH_hist} shows results for a single realization of a 4-year observation time, Table~\ref{tbl:MBH_stats_fin_res} contains averaged statistics for the entire catalogs described in Ref.~\cite{Barausse:2023yrx}, which collectively correspond to 100 years of MBH data: for instance, to determine the number of events with a detectable quadratic mode in 4 years, we divide the total number of events with $\rho_{22\times22}>8$ by 25.

\begin{table*}
\caption{Averaged statistics on the MBH binaries observed by LISA. The first and second columns show the total number of mergers and the number of events with observable IMR expected in a 4-year mission lifetime for each catalog. The third and fourth columns show the same quantities when we implement the mass ratio cutoff ($q<10$). The fifth and sixth 
columns list the number of events having SNR above threshold 
for the dominant $(22)$ linear QNM and for the $(22 \times22)$ quadratic QNM (for the $q<10$ events only). In the last two columns we list the average and maximum SNRs of the $(22 \times22)$ mode (again, for $q<10$ events). Numbers without and with parentheses represent values for the finite-resolution and extrapolated models, respectively.}
\begin{ruledtabular}\label{tbl:MBH_stats_fin_res} 
\addtolength{\tabcolsep}{-1.7em}
\begin{tabular}{lllllllll}
 & Events in & Num. with  & Events in 4 & Num. with  & Num. with & Num. with  & Mean  & Max \\ 
 &  4 yrs &  $\rho_{\rm{IMR}}>8$ & yrs  ($q<10$) & $\rho_{\rm{IMR}}>8$  ($q<10$) & $\rho_{22}>8$ & $\rho_{22\times22}>8$ & $\rho_{22\times22}$  &  $\rho_{22\times22}$ \\\hline
 
 HS-nod-noSN (B+20) & 16288(39785) & 16284(39764) & 11978(29383) & 11977(29380) & 6704(20951) & 1098(5623) & 3(5) & 905(2211)\\
 
 LS-nod-noSN (B+20) & 1313(1672) & 224(271) & 1193(1529) & 132(163) & 11(13) & 3(4) & 0.3(0.3) & 1149(1152) \\
 
 LS-nod-SN (B+20) & 1279(1626)& 6(7) & 1276(1622) & 5(6) & 0(6) & 0(0) & 0(0) & 94(418) \\
 
 pop-III-d(K+16) & 689(1430) & 206(382) &  662(1376) & 180(334) & 5(15) & 2(7) & 0.6(0.7) & 1725(1024) \\
  
 Q3-nod (K+16) & 470(660) & 470(659) & 359(516) & 359(516) &  277(427) & 77(139) & 8(14) & 964(1744) \\
 
 Q3-d (K+16) & 33(74) & 33(74) & 31(70) & 31(70) & 28(66) & 22(55) & 74(93) & 2194(3870) \\
\end{tabular}
\end{ruledtabular}
\end{table*}

\begin{figure*}[ht!]
\includegraphics[width=0.95
\textwidth]{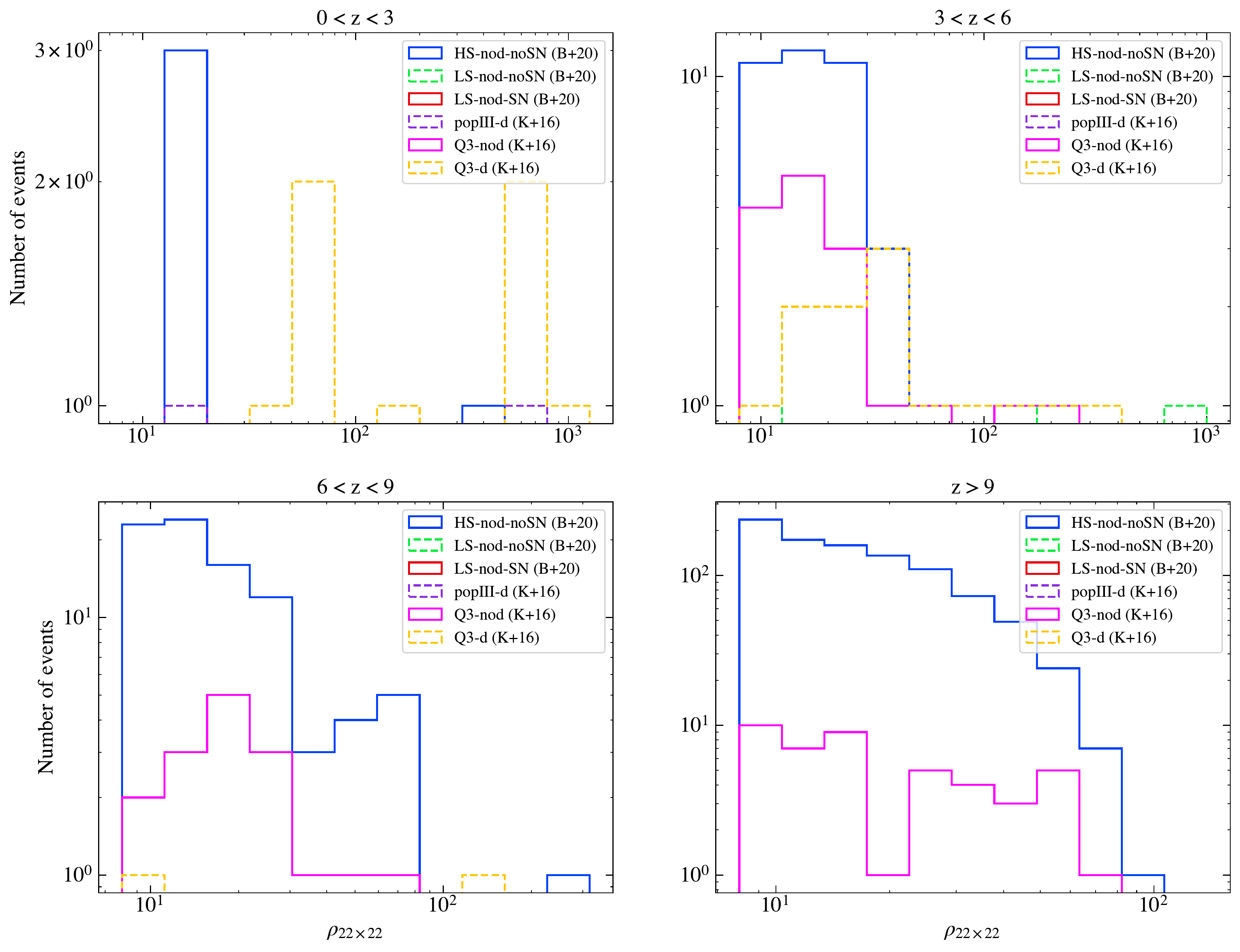}
\caption{Same as Fig.~\ref{fig:MBH_hist}, but with the quadratic mode SNR distributions further broken down by redshift range. The plots refer to the same finite-resolution catalogs shown in Fig.~\ref{fig:MBH_hist}.}
\label{fig:MBH_hist_sliced}
\end{figure*}

While there is large variation between different astrophysical MBH models, the number of events with a detectable quadratic mode can be several orders of magnitude larger than what we found for ground-based observations using \texttt{models~I-III}. Although we do not show all of our results here for brevity, we consistently find $\mathcal{O}(10-10^3)$ events per 4 years with $\rho_{22\times22}>8$ for all models with heavy seeds (including the other models in Ref.~\cite{Barausse:2023yrx}, i.e., ``HS-nod-SN (B+20)'' and ``HS-nod (B12)'').
In contrast, prospects for detecting the ($22\times22$) mode are no better (in fact, they are somewhat worse) for models with light seeds, compared to the stellar mass and IMBH populations. (See Appendix~\ref{app:table_II_illustration} for a visualization of the comparative number of events with detectable quadratic mode for heavy seed vs. light seed MBH models.)

To further understand the conditions under which we will be more likely to detect the ($22\times22$) mode, we also divide the results plotted in Fig.~\ref{fig:MBH_hist} into subsets of data spanning different redshift ranges. The results are shown in Fig.~\ref{fig:MBH_hist_sliced}. Perhaps counterintuitively, by far the largest number of MBH events with a detectable quadratic mode are at a redshift greater than 9. This can be understood in terms of gravitational redshift (compare e.g. Fig.~1 of Ref.~\cite{Baibhav:2018rfk}): MBHs with source-frame masses of $\mathcal{O}(10^5-10^6)M_\odot$, which are the average masses of the ``HS'' and ``Q3'' models, have {\em redshifted} masses that fall well within the LISA noise band. 

To conclude, our SNR calculations suggest that MBHs with source frame masses of $\sim10^5-10^6 M_\odot$ and $z\gtrsim9$ will be excellent candidates for detecting the ($22\times22$) nonlinear mode with LISA. 

\begin{figure*}[ht!]  \includegraphics[width=0.7\textwidth]{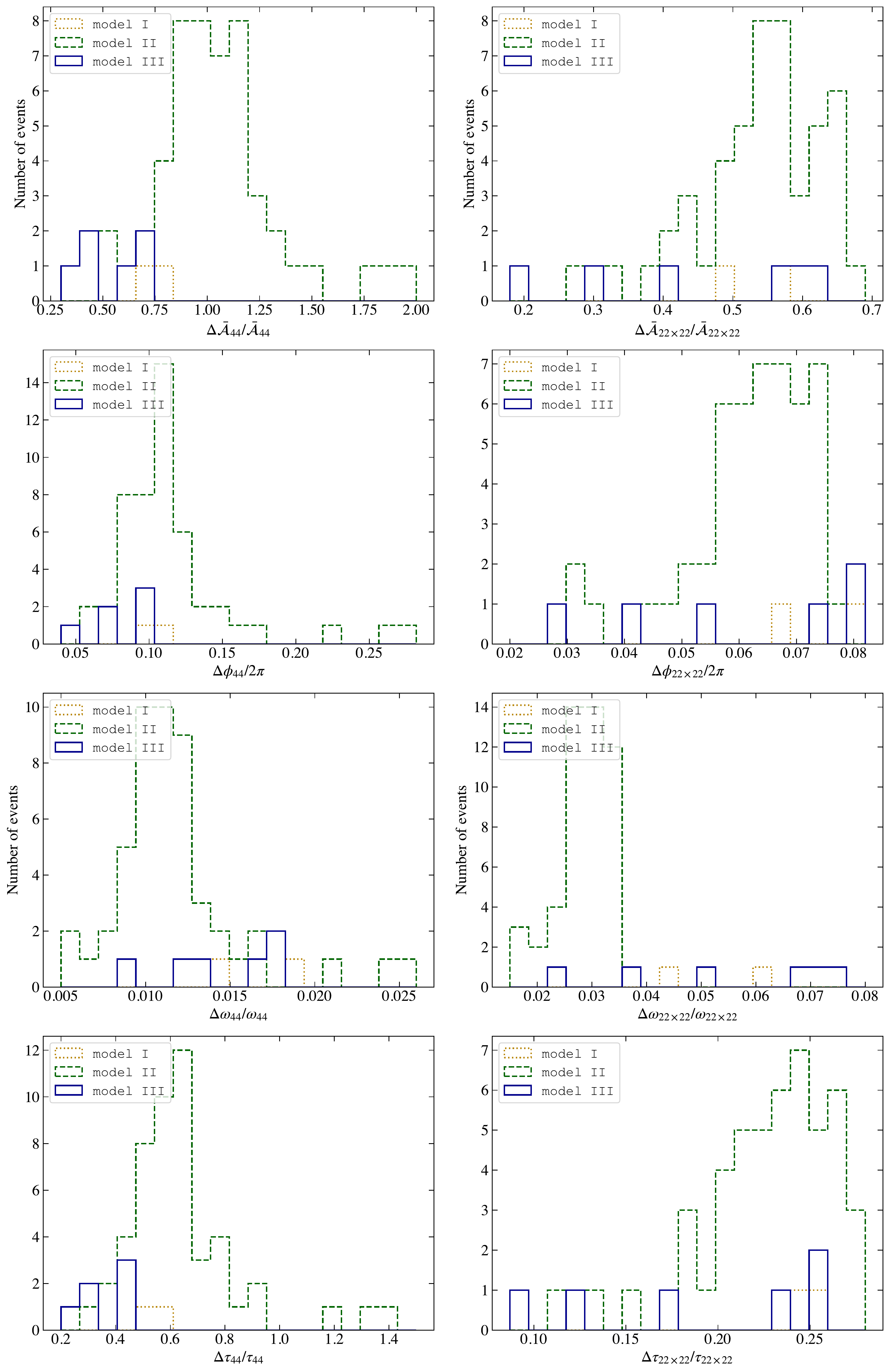}
  \caption{Histograms of the errors on the QNM parameters (amplitude, phase, frequency and damping time, from top to bottom) for the subset of \texttt{model~I-III} events such that the quadratic mode is observable by ET. Results for CE are qualitatively similar.}
\label{fig:errors_ground_based_ET}
\end{figure*}

\section{Parameter estimation of nonlinear modes}
\label{sec:measurability}

The SNR results presented so far are a simple figure of merit to assess the detectability of nonlinear modes with future ground-based and space-based detectors. However, a more complete understanding of our ability to distinguish these quadratic QNMs from ``ordinary'' linear QNMs requires a parameter estimation analysis, and it should take into account correlations among the waveform parameters.
Here we assess parameter estimation accuracy for quadratic modes, as observed by both ground-based and space-based detectors, through the FIM formalism. We use the same conventions adopted for the SNR calculations: in particular,  we use a starting time $t_0=12M_f$, and we discard all events with mass ratio $q>10$.

As highlighted in Sec.~\ref{sec:fisher_intro}, for our purposes it is sufficient to compute the $8\times8$ block diagonal elements of the Fisher matrix including the $(44)$ and the ($22\times22$) QNMs, because these errors are independent from the errors corresponding to QNMs with $(\ell m)\neq(44)$.

\begin{figure*}[ht!]
\includegraphics[width=0.7\textwidth]{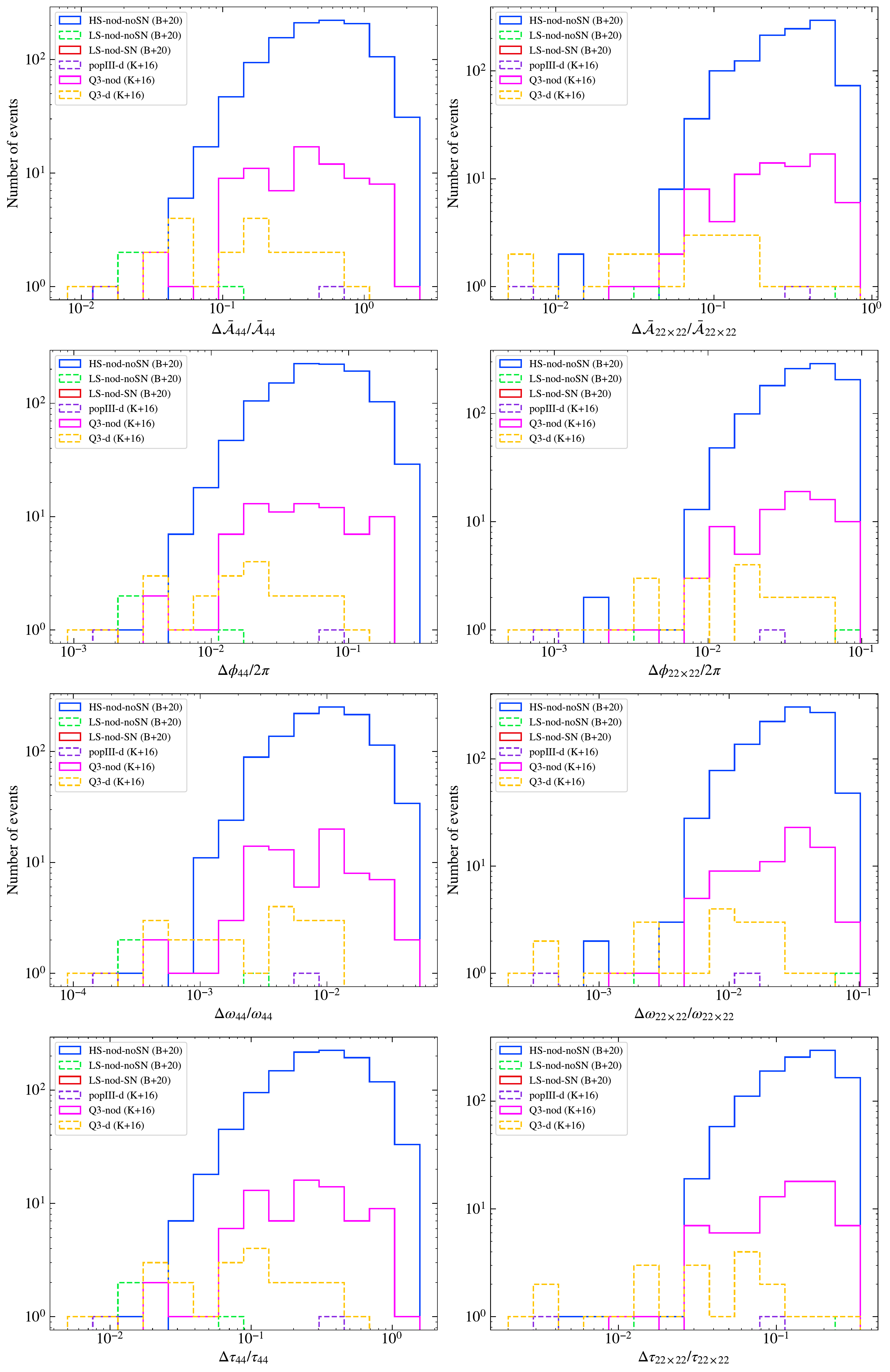}
\caption{Histograms of QNM parameter errors for the events in the MBH catalogs with a detectable quadratic mode SNR. }
\label{fig:errors_LISA}
\end{figure*}

\subsection{Error estimates with ground-based XG detectors}

In Fig.~\ref{fig:errors_ground_based_ET} we plot histograms of the errors on the $(44)$ and $(22\times22)$ amplitude, phase, frequency, and damping time, focusing on the \texttt{model~I-III} events
such that the SNR of the quadratic mode $\rho_{22\times22}>8$, as observed by ET (see bottom panel of Fig.~\ref{fig:hist_rapster_and_parspec}).  The results for CE, not shown here for brevity, are qualitatively similar.

The posterior distributions on the parameters of the quadratic mode are all informative. The frequency of the $(22\times 22)$ QNM is always measured with an accuracy of order $10\%$ or better, with relative errors clustering around $3\%$ in the most optimistic case (\texttt{model II}).
The frequency of the dominant $(44)$ linear QNM is measured even better.
The quadratic mode damping time is also measured quite well, with relative errors generally smaller than $\sim 30\%$.
Interestingly, for the majority of the events, the errors on the parameters of the quadratic $(22\times22)$ QNM are comparable to, or sometimes even smaller than, those of the linear $(44)$ QNM. This is because the amplitude of the quadratic mode is typically larger than the amplitude of the $(44)$ linear QNM at $t_0=12M_f$.
In particular, this explains why the relative error on the damping time of the $(44)$ linear QNM is slightly worse than the corresponding error on the damping time of the $(22\times 22)$ QNM.
Note also that the posteriors of the quadratic QNM amplitudes do not show support for $\Bar{\mathcal{A}}_{22\times 22}=0$.

Let us summarize our expected ability to measure quadratic QNM parameters with ground-based detectors in the most optimistic case.

Focusing on the model with the largest number of events satisfying $\rho_{22\times22}>8$ (\texttt{model~II}, with 50 events satisfying this criterion), we find 14  events with relative errors on the quadratic mode amplitude smaller than 50\%, as measured with ET. The $(22\times22)$ mode phases, frequencies and damping times are measured much better: all 50 events have $\Delta \phi_{22\times22}/2\pi < 8\%$, relative errors on $\omega_{22\times22}$ smaller than $4\%$, and relative errors on $\tau_{22\times22}$ smaller than $30\%$. For CE, we have 4 out of the 10 detectable events with $\Delta \Bar{\mathcal{A}}_{22\times22}/\Bar{\mathcal{A}}_{22\times22}<50\%$; again, all detectable events have relative errors on $\phi_{22\times22} < 8\%$, relative errors on $\omega_{22\times22} < 4\%$, and  relative errors on $\tau_{22\times22} < 30\%$.

In the most pessimistic case, i.e., \texttt{model~I}, we find only 2 events with $\rho_{22\times22}>8$ as measured by ET, and no events with a detectable quadratic mode as measured by CE. The errors for the 2 events measured with ET are similar to the errors for \texttt{model~II}.

\subsection{Error estimates with LISA}

In Fig.~\ref{fig:errors_LISA} we plot the estimated errors on the $(44)$ and ($22\times22$) mode parameters for the MBH events with $\rho_{22\times22}>8$. 
Many of the trends are similar to those observed in Fig.~\ref{fig:errors_ground_based_ET}. For example, the ($22\times22$) damping time, amplitude and phase are generally measured somewhat better than the corresponding $(44)$ parameters, and the real parts of the QNM frequencies are measured quite well, with typical errors of the order of a few percent.
The major difference, of course, is that the number of MBH events with such small errors is much larger than the number of events originating from the stellar mass and IMBH mergers observable by ground-based detectors.

In the most optimistic case (i.e., for the ``HS-nod-noSN (B+20)'' model, with 1098 events satisfying $\rho_{22\times22}>8$), we find 896 events with relative error on the quadratic mode amplitude less than 50\%. All 1098 events have relative error on $\phi_{22\times22}$ and $\omega_{22\times22}$ less than 10\%, and relative error on $\tau_{22\times22}$ less than 30\%.
These results refer to the finite-resolution catalogs, which, as explained in Sec.~\ref{sec:MBH_description}, provide a lower limit to our predicted detectable quadratic QNM rates. 
Note also that the posteriors of the quadratic QNM amplitudes show no support for $\Bar{\mathcal{A}}_{22\times22}=0$ for all of the MBH models we consider.
In the most pessimistic case (``LS-nod-SN (B+20)''), we find no events with a detectable quadratic mode. 

\section{Conclusions}
\label{sec:conclusions}

In this paper we have built on recent developments in the construction of amplitude and phase fits for nonlinear quasinormal modes~\cite{Cheung:2023vki} to forecast the detectability of the $(22\times22)$ quadratic mode with XG gravitational wave detectors. We have estimated the SNR of the quadratic mode for several binary BH population models that may be observed by ground-based and space-based detectors in the upcoming decades. Our main finding is that a 15~km ET could detect up to a few tens of events per year with an observable quadratic mode, and CE has similar prospects for observing quadratic modes. In contrast, the most optimistic astrophysical MBH formation models allow for the possibility that LISA may observe up to $\mathcal{O}(1000)$ events with a detectable quadratic mode in its nominal 4-year observation time.

For the subset of events with a detectable $(22\times22)$  mode, we used a FIM analysis to estimate fairly small measurement errors on the quadratic mode amplitude, phase, and frequencies. It will be important to validate these preliminary findings with Bayesian parameter estimation codes such as \texttt{PyRing}~\cite{Carullo:2019flw,pyRing} and \texttt{Ringdown}~\cite{Isi:2021iql,IsiRingdown}, or neural posterior estimation methods~\cite{Crisostomi:2023tle}.

In this study, we have only investigated the ability of future gravitational wave interferometers to detect the quadratic $(22\times22)$ mode sourced by the square of the fundamental $(22)$ mode, simply because this mode is the easiest to identify in numerical relativity simulations. It will be interesting to further investigate the detectability of other nonlinearities in the merger signal, including (but not limited to) subdominant quadratic QNMs.

\begin{acknowledgments}
We thank Mesut Çalışkan for providing a \texttt{python} script with which to make the plots in Appendix~\ref{app:table_II_illustration}.
  S.~Yi  is supported by the NSF Graduate Research Fellowship Program under Grant No. DGE2139757.
  M.~Ho-Yeuk Cheung is a Croucher Scholar supported by the Croucher Foundation.
  K.~Kritos is supported by the Onassis Foundation - Scholarship ID: F ZT 041-1/2023-2024.
  S.~Yi, M.~H.-Y. Cheung, K.~Kritos, and E.~Berti are supported by NSF Grants No. AST-2307146, PHY-2207502, PHY-090003 and PHY-20043, by NASA Grants No. 20-LPS20-0011 and 21-ATP21-0010, by the John Templeton Foundation Grant 62840, by the Simons Foundation.
  E.~Berti and A.~Maselli are supported by the Italian Ministry of Foreign Affairs and International Cooperation grant No.~PGR01167.
  A.~Maselli acknowledges financial support from MUR PRIN Grant No. 2022-Z9X4XS, funded by the European 
  Union - Next Generation EU.
  This work was carried out at the Advanced Research Computing at Hopkins (ARCH) core facility (\url{rockfish.jhu.edu}), which is supported by the NSF Grant No.~OAC-1920103.
  E.~Barausse and A.~Kuntz acknowledge support from the European Union’s H2020 ERC Consolidator Grant ``GRavity from Astrophysical to Microscopic Scales'' (Grant No. GRAMS-815673), the PRIN 2022 grant ``GUVIRP - Gravity tests in the UltraViolet and InfraRed with Pulsar timing'', and the EU Horizon 2020 Research and Innovation Programme under the Marie Sklodowska-Curie Grant Agreement No. 101007855.
\end{acknowledgments}

\appendix 

\section{SNR and parameter estimation results at different QNM starting times}
\label{app:bracket}

In this appendix, we argue that despite the non-negligible dependence of the SNR and parameter errors on the choice of $t_0$, our use of $t_0=12M_f$ for the main results nevertheless provides a useful estimate of the quadratic mode detectability with XG observatories. 

To this end, we first take a subset of the events plotted in Figs.~\ref{fig:MBH_hist} and~\ref{fig:MBH_hist_sliced} having the ``loudest'' quadratic mode SNRs ($\rho_{22\times 22}>50$ at $t_0=12M_f$), and we recompute their SNR at $t_0=7M_f$ and $t_0=17M_f$. The results are shown in Fig.~\ref{fig:MBH_t0_test}, where the SNR values computed at $t_0=7M_f$ ($t_0=17M_f$) are plotted with dashed (dotted) lines, and the results at $t_0=12M_f$ shown in the main text are plotted as solid lines for reference. For simplicity we only focus on the loudest events in the ``HS-nod-noSN (B+20)'' catalog, but we have observed similar trends for the loudest events in other MBH catalogs. 

\begin{figure}[t]
\includegraphics[width=0.45\textwidth]{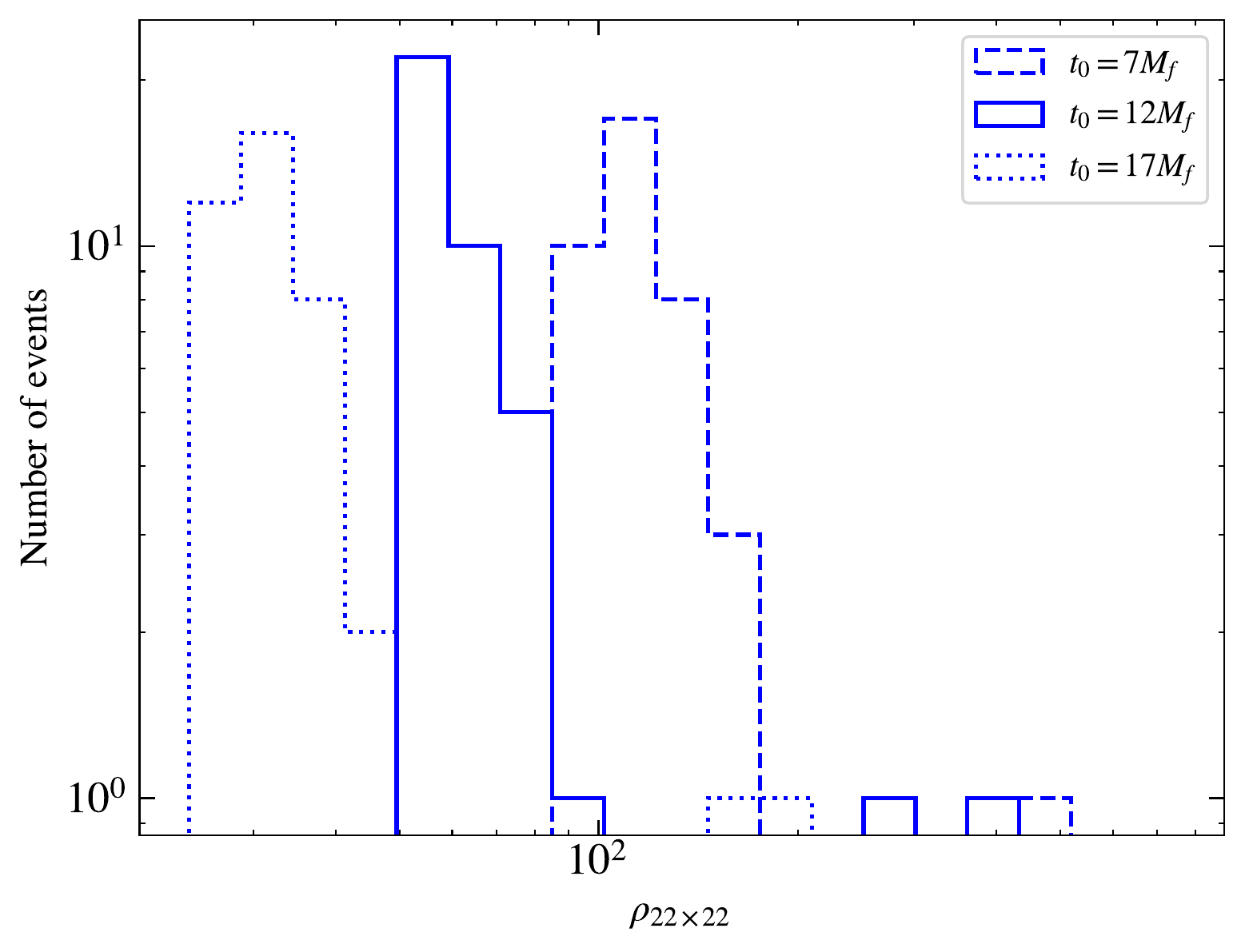}
\caption{Quadratic mode SNR of the ``loudest'' MBH events in the ``HS-nod-noSN (B+20)'' model, calculated at different starting times, $t_0$. Dashed, solid, and dotted lines correspond to SNRs calculated at $t_0=7,12,$ and $17M_f$, respectively.}
\label{fig:MBH_t0_test}
\end{figure}

As expected, the SNRs calculated at earlier (later) times are larger (smaller) than the SNRs computed at the fiducial time $t_0=12M_f$ that we use throughout the main text. However, the results change by less than one order of magnitude: the minimum quadratic mode SNR of all events in Fig.~\ref{fig:MBH_t0_test} is 24.3 at $t_0=17M_f$ and 90.9 at $t_0=7M_f$, and the maximum SNR is 177.9 at $t_0=17M_f$ and 745.5 at $t_0=7M_f$. Therefore, when we can claim detection of the quadratic mode fairly confidently, increasing or decreasing $t_0$ by $\sim 5M_f$ is not usually enough to withdraw a detection claim. 

In Fig.~\ref{fig:ground_to_test} we perform a similar test for the detectable events in the \texttt{model~II} catalog as observed by ET. Here, we see that since many of the events were only marginally detectable at $t_0=12M_f$, increasing $t_0$ by $5M_f$ will cause the quadratic mode SNR of some of the events to fall below our detectability threshold ($\rho_{22\times22}=8$, marked by the vertical red line). On the bright side, if $t_0$ is smaller than our fiducial value by a few $M_f$, many of the events with marginally detectable SNRs (very close to $\rho_{22\times22}=8$) at $t_0=12M_f$ are louder. 

\begin{figure}[t]
\includegraphics[width=0.42\textwidth]{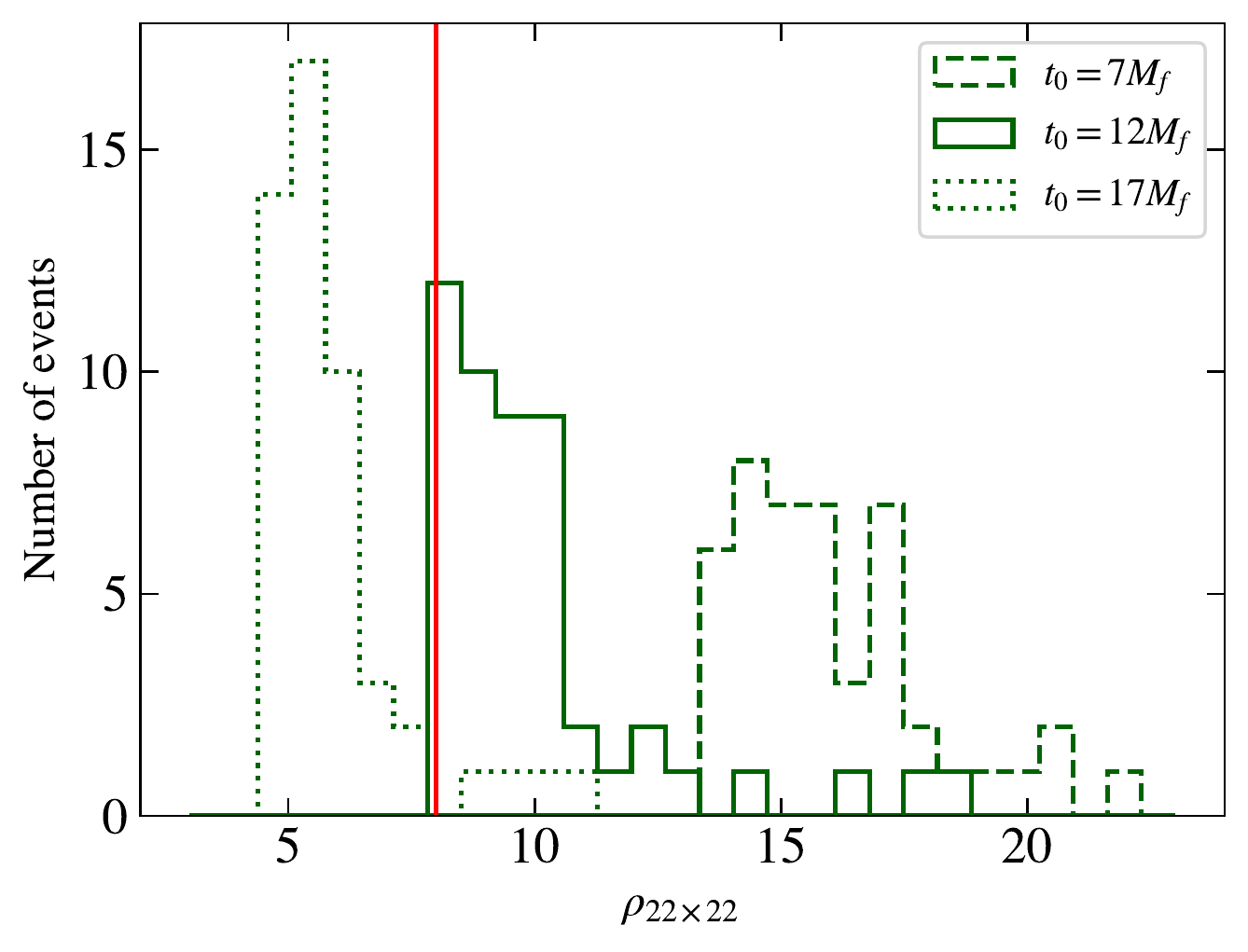}
\caption{Quadratic mode SNRs of events in the \texttt{model~II} catalog classified as ``detectable'' in the main text ($\rho_{22\times22}>8$ at $t_0=12M_f$), calculated at different starting times. Dashed, solid, and dotted lines correspond to SNRs calculated at $t_0=7,12,$ and $17M_f$, respectively. The vertical red line marks $\rho_{22\times22}=8$. }
\label{fig:ground_to_test}
\end{figure}

Finally, in Fig.~\ref{fig:MBH_errors_t0_test} we plot the errors on the $(22\times22)$ mode parameters calculated at different values of $t_0$ for the same ``loud'' MBH events shown in Fig.~\ref{fig:MBH_t0_test}. While the errors do increase (decrease) for larger (smaller) values of $t_0$, we see that the shifts in the distributions are fairly moderate. In particular, the relative errors on all mode parameters for these ``loudest'' events remain smaller than unity even at a pessimistic value of $t_0=17M_f$.

\begin{figure}[t]
\includegraphics[width=0.48\textwidth]{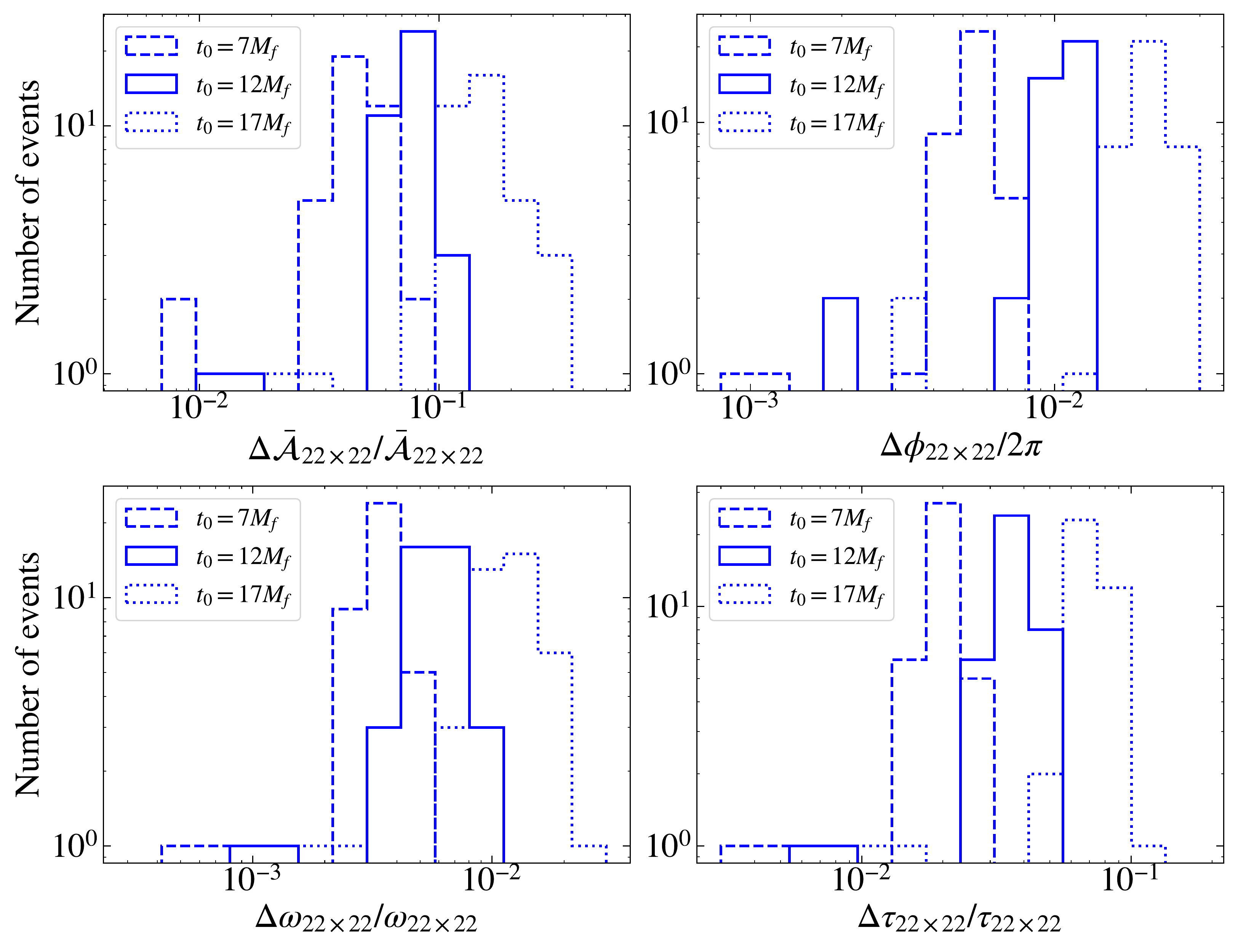}
\caption{Errors calculated at different values of $t_0$ for the ``loudest'' MBH events in the ``HS-nod-noSN (B+20)'' model. Once again, dashed, solid, and dotted lines correspond to errors calculated at $t_0=7,12,$ and $17M_f$, respectively.}
\label{fig:MBH_errors_t0_test}
\end{figure}

\section{Merger rate of binary BHs from star clusters}\label{app:merger_rate}

In this appendix, we describe how to properly reweigh the {\tt Rapster} simulation output and compute the merger rate from our generated set of events.
Our {\tt Rapster} simulations have been carried out in a fixed comoving volume $\Delta V_{\rm c}$.
We define the comoving source-frame merger rate density to be ${\cal R}_{\rm s}=dN_{\rm me}/(dV_{\rm c}dt_{\rm s})$.
To compute this quantity we count the number of binary BH mergers within $[z,z+\Delta z]$, denoted by $\Delta N_{\rm me,simul}^{[z,z+\Delta z]}$.
Each such event ``$i$'' has been produced by a cluster that formed at a larger redshift $z_{\rm cl}^{(i)}$.
Since we have simulated $\Delta N_{\rm cl,simul}^{[z_{\rm cl}^{(i)}, z_{\rm cl}^{(i)}+\Delta z]}$ clusters, we need to correct for the true number of clusters that form at $z_{\rm cl}^{(i)}$, that is,
\begin{align}
    \Delta N_{\rm cl,true}^{[z_{\rm cl}^{(i)},z_{\rm cl}^{(i)}+\Delta z]} = \psi_{\rm cl}(z_{\rm cl}^{(i)}) \Delta V_{\rm c}\Delta t_{\rm s}(z_{\rm cl}^{(i)})\,,
\end{align}
where $\psi_{\rm cl}$ is the comoving source-frame cluster formation rate density given by Eq.~(11) of Ref.~\cite{Mapelli:2021gyv}.
As such, the true number of mergers in $[z,z+\Delta z]$ becomes
\begin{align}
    \Delta N_{\rm me,true}^{[z,z+\Delta z]} = \sum_{i=1} ^ {\Delta N_{\rm me,simul}^{[z,z+\Delta z]}} {\Delta N_{\rm cl,true}^{[z_{\rm cl}^{(i)},z_{\rm cl}^{(i)}+\Delta z]} \over \Delta N_{\rm cl,simul}^{[z_{\rm cl}^{(i)},z_{\rm cl}^{(i)}+\Delta z]}}\,.
\end{align}
Thus, the source-frame merger rate per unit comoving volume can be approximated as
\begin{equation}
    \mathcal{R}_{\rm s}(z)\simeq\sum^{\Delta N_{\rm me, simul}^{[z,z+\Delta z]}}_{i=1} \frac{\psi(z_{\rm cl}^{(i)})}{N_{\rm cl,simul}^{[z_{\rm cl}^{(i)},z_{\rm cl}^{(i)}+\Delta z]}} \frac{\Delta t_{\rm s}(z_{\rm cl}^{(i)})}{\Delta t_{\rm s}(z_{\rm me}^{(i)})}\,.
    \label{eq:Rs(z)_approximation}
\end{equation}

By rearranging the definition of the observer-frame volumetric merger rate ${\cal R}_{\rm o} = {\cal R}_{\rm s} {dt_{\rm o} / dt_{\rm s}}$, where $dt_{\rm o}/dt_{\rm s}=(1+z)$, we obtain the cumulative number of mergers within redshift $z$ per year [see also Eq.~(17) from~\cite{Rodriguez:2016kxx}]:
\begin{align}
    R_{\rm o}(<z) = \int_0^{z} {dz'\over 1+z'} {\cal R}_{\rm s}(z') {dV_{\rm c}\over dz'}.
    \label{eq:cumulative_merger_rate}
\end{align}
Integrating out to a redshift of 10 results in $\simeq2\times10^4$ globular cluster mergers per year anywhere in the Universe. 

When we sample $2\times10^4$ events from the \texttt{Rapster} catalog with a probability density given by the integrand of Eq.~\eqref{eq:cumulative_merger_rate}, we find no events with $\rho_{22\times22}>8$.
Nevertheless, we have reasons to believe that the actual number of these mergers per year may be somewhat larger. 
First, the \texttt{Rapster} code likely underestimates the merger rate per cluster by a factor of a few (see Fig.~5 of~\cite{Kritos:2022ggc}).
Moreover, dynamically formed binary BHs may also originate from young massive and nuclear star clusters, in addition to the globular cluster contribution~\cite{Mapelli:2021gyv}.
Even the number density of globular clusters is uncertain, leading to a large merger rate uncertainty (see Fig.~3 of~\cite{Mandel:2021smh}), with an upper limit that is consistent with the currently inferred value from gravitational wave observations~\cite{KAGRA:2021duu}.
We roughly estimate these considerations to collectively increase the number of mergers from dynamical channels by a factor of a few per year.
Altogether, then, to generate the catalog of {\tt model III} used for our analysis, we sample $10^5$ events with a probability density given by the integrand of Eq.~\eqref{eq:cumulative_merger_rate}.

\section{SNR and Fisher matrix expressions for multiple modes in the same angular harmonic}\label{app:orthogonality}

Consider the complex spherical harmonic functions defined in Eq.~\eqref{eqn:yplus_ycross}. The following averaged products are zero for any combination of modes with $(\ell m)=(22),(21),(33),(44)$:
\begin{eqnarray}
        \braket{(\hat{Y}^{\ell m}_+)^2}=\braket{(\hat{Y}^{\ell m *}_+)^2}&=&0\,,\\
        \braket{(\hat{Y}^{\ell m}_\times)^2}=\braket{(\hat{Y}^{\ell m*}_\times)^2}&=&0\,,\\
        \braket{(\hat{Y}^{\ell m}_+)(\hat{Y}^{\ell m}_\times)}=\braket{(\hat{Y}^{\ell m}_+)^*(\hat{Y}^{\ell m}_\times)^*}&=&0\,,\\
        \braket{(\hat{Y}^{\ell m}_+)(\hat{Y}^{\ell m}_\times)^*}=\braket{(\hat{Y}^{\ell m}_+)^*(\hat{Y}^{\ell m}_\times)}&=&0\,.
\end{eqnarray}
The products
\begin{eqnarray}
    \braket{(\hat{Y}^{\ell m}_+)(\hat{Y}^{\ell m}_+)^*}=\braket{|\hat{Y}^{\ell m}_+|^2}\\
    \braket{(\hat{Y}^{\ell m}_\times)(\hat{Y}^{\ell m}_\times)^*}=\braket{|\hat{Y}^{\ell m}_\times|^2}
\end{eqnarray}
are not always zero; however, the sum 
\begin{eqnarray}
    \braket{|\hat{Y}^{\ell m}_+|^2+|\hat{Y}^{\ell m}_\times|^2}
\end{eqnarray}
is equal to zero for any two modes with different indices $(\ell m)$, and equal to $1/\pi$ for two modes with the same $(\ell m)$. 

These relations have consequences for how we compute the SNR and Fisher matrices for multiple modes. Specifically, when treating modes within the same $(\ell m)$ harmonic, we cannot simply assume a block-diagonal Fisher matrix, or that the SNR is given by a sum in quadrature of the individual mode SNRs. 

\subsection{Modifications to the SNR for two modes with the same $(\ell m)$}

The frequency-domain ringdown waveform is given by Eqs.~\eqref{eq:tildehp}-\eqref{eqn:waveform_full}.
If we want to compute the SNR of a waveform with just two modes (i.e., $\Tilde{h}_+(f)=\Tilde{h}_+^{(1)}+\Tilde{h}_+^{(2)}$ and $\Tilde{h}_\times(f)=\Tilde{h}_\times^{(1)}+\Tilde{h}_\times^{(2)}$), 
then using $\braket{F^2_{+,\times}}=\frac{1}{5}$ and $\braket{F_+F_\times}=0$, we have, after averaging over all angles,
\begin{eqnarray}
    \braket{\Tilde{h}(f)\Tilde{h}^*(f)}=\frac{1}{5}[ |\Tilde{h}_+^{(1)}|^2+\Tilde{h}_+^{(1)}\Tilde{h}_+^{(2)*}+\Tilde{h}_+^{(1)*}\Tilde{h}_+^{(2)}+|\Tilde{h}_+^{(2)}|^2\nonumber\\+|\Tilde{h}_\times^{(1)}|^2+\Tilde{h}_\times^{(1)}\Tilde{h}_\times^{(2)*}+\Tilde{h}_\times^{(1)*}\Tilde{h}_\times^{(2)}+|\Tilde{h}_\times^{(2)}|^2]\,. \nonumber \\
\end{eqnarray}

If the modes have different indices $(\ell m)$, then by the orthogonality of the spherical functions, this reduces to 
\begin{eqnarray}
    \braket{\Tilde{h}(f) )\Tilde{h}^*(f)}=\frac{1}{5}\left[|\Tilde{h}_+^{(1)}|^2+|\Tilde{h}_+^{(2)}|^2+|\Tilde{h}_\times^{(1)}|^2+|\Tilde{h}_\times^{(2)}|^2\right]\nonumber\\
    =\frac{\Bar{\mathcal{A}}^2_1}{10\pi}\left( (b_+^{(1)})^2 + (b_-^{(1)})^2\right)+\frac{\Bar{\mathcal{A}}^2_2}{10\pi}\left( (b_+^{(2)})^2 + (b_-^{(2)})^2\right)\,.\nonumber\\
\end{eqnarray}
If the two modes have the same $(\ell m)$, we have additional terms from the following nonzero quantities:
\begin{widetext}
\begin{eqnarray}
    \Tilde{h}_+^{(1)}\Tilde{h}_+^{(2)*}=\frac{\Bar{\mathcal{A}}_1\Bar{\mathcal{A}}_2}{2}\left( b_+^{(1)}b_+^{(2)} e^{i(\phi_1-\phi_2)} \hat{Y}_{+}^{(1)}\hat{Y}_{+}^{(2)*}+ b_-^{(1)}b_-^{(2)} e^{-i(\phi_1-\phi_2)}\hat{Y}_{+}^{(1)*}\hat{Y}_{+}^{(2)}\right)\,,\\
    \Tilde{h}_+^{(1)*}\Tilde{h}_+^{(2)}=\frac{\Bar{\mathcal{A}}_1\Bar{\mathcal{A}}_2}{2}\left( b_+^{(1)}b_+^{(2)} e^{-i(\phi_1-\phi_2)} \hat{Y}_{+}^{(1)*}\hat{Y}_{+}^{(2)}+ b_-^{(1)}b_-^{(2)} e^{i(\phi_1-\phi_2)}\hat{Y}_{+}^{(1)}\hat{Y}_{+}^{(2)*}\right) ,\\
    \Tilde{h}_\times^{(1)}\Tilde{h}_\times^{(2)*}=\frac{\Bar{\mathcal{A}}_1\Bar{\mathcal{A}}_2}{2}\left( b_+^{(1)}b_+^{(2)} e^{i(\phi_1-\phi_2)} \hat{Y}_{\times}^{(1)}\hat{Y}_{\times}^{(2)*}+b_-^{(1)}b_-^{(2)} e^{-i(\phi_1-\phi_2)}\hat{Y}_{\times}^{(1)*}\hat{Y}_{\times}^{(2)}\right) ,\\
    \Tilde{h}_\times^{(1)*}\Tilde{h}_\times^{(2)}=\frac{\Bar{\mathcal{A}}_1\Bar{\mathcal{A}}_2}{2}\left( b_+^{(1)}b_+^{(2)} e^{-i(\phi_1-\phi_2)} \hat{Y}_{\times}^{(1)*}\hat{Y}_{\times}^{(2)}+b_-^{(1)}b_-^{(2)} e^{i(\phi_1-\phi_2)}\hat{Y}_{\times}^{(1)}\hat{Y}_{\times}^{(2)*}\right) .
\end{eqnarray}
\end{widetext}
Adding the first and third lines, and the second and fourth lines, and recalling that $\braket{|\hat{Y}^{\ell m}_+|^2+|\hat{Y}^{\ell m}_\times|^2}=\frac{1}{\pi}$ for two modes with the same $(\ell m)$, the term we must add to the SNR for two modes in the same harmonic is
\begin{eqnarray}
   &&\frac{1}{5}\left[\Tilde{h}_+^{(1)}\Tilde{h}_+^{(2)*}+\Tilde{h}_+^{(1)*}\Tilde{h}_+^{(2)}+\Tilde{h}_\times^{(1)}\Tilde{h}_\times^{(2)*}+ \Tilde{h}_\times^{(1)*}\Tilde{h}_\times^{(2)}\right]\nonumber \\
   &&\;\;\;\;\;\;=\frac{\Bar{\mathcal{A}}_1\Bar{\mathcal{A}}_2}{10\pi} \left( b_+^{(1)}b_+^{(2)} e^{i(\phi_1-\phi_2)} + b_-^{(1)}b_-^{(2)} e^{-i(\phi_1-\phi_2)}\right)\nonumber\\
&&\;\;\;\;\;\;+\frac{\Bar{\mathcal{A}}_1\Bar{\mathcal{A}}_2}{10\pi} \left( b_+^{(1)}b_+^{(2)} e^{-i(\phi_1-\phi_2)} + b_-^{(1)}b_-^{(2)} e^{i(\phi_1-\phi_2)}\right)\nonumber\\
   &&=\frac{\Bar{\mathcal{A}}_1\Bar{\mathcal{A}}_2}{5\pi}\left[(b_+^{(1)}b_+^{(2)} +b_-^{(1)}b_-^{(2)})\mathrm{cos}(\phi_1-\phi_2)\right]
\end{eqnarray}
after we average over the spherical functions. 

\subsection{Modifications to FIM expressions for two modes with the same values of $(\ell m)$}
The top left and bottom right $4\times4$ blocks of the total $8\times8$ FIM for two QNMs with identical $(\ell m)$ indices are given by the expressions in Appendix~D of~\cite{Maselli:2023khq}. 

In addition, we have the following nonzero components on the off-diagonal blocks of the FIM. We denote the 8 mode parameters by $\Vec{\theta}=(\Bar{\mathcal{A}}_1,\phi_1,\omega_1,\tau_1,\Bar{\mathcal{A}}_2,\phi_2,\omega_2,\tau_2)$; in this paper, we have primarily been considering the indices 1 and 2 to stand for mode indices $(44)$ and $(22\times22)$. 
What we call `` $\partial_{\theta_i} h  \left(\partial_{\theta_j} h \right)^*$'' here corresponds to the expressions labeled ``$\Gamma_{\theta_i \theta_j}$'' in Appendix~D of Ref.~\cite{Maselli:2023khq}.
\begin{widetext}
\begin{eqnarray}
    \partial_{\Bar{\mathcal{A}}_1} h \, (\partial_{\Bar{\mathcal{A}}_2}h)^* &=& \frac{1}{10 \pi }\left[e^{i (\phi_2-\phi_1)} b_-^{(1)} b_-^{(2)}+e^{i(\phi_1-\phi_2)} b_+^{(1)} b_+^{(2)} \right]\\
    \partial_{\Bar{\mathcal{A}}_1} h \, (\partial_{\phi_2}h)^* &=& \frac{i \Bar{\mathcal{A}}_2}{10 \pi } \left[e^{i (\phi_2-\phi_1)} b_-^{(1)} b_-^{(2)}- e^{i(\phi_1-\phi_2)} b_+^{(1)} b_+^{(2)}\right]\\
    \partial_{\Bar{\mathcal{A}}_1} h \, (\partial_{\omega_2}h)^* &=& \frac{\Bar{\mathcal{A}}_2}{10 \pi } \left[e^{i (\phi_2-\phi_1)} b_-^{(1)} b_{-,\omega_2}^{(2)}+e^{i(\phi_1-\phi_2)} b_+^{(1)} b_{+,\omega_2}^{(2)}\right]\\
    \partial_{\Bar{\mathcal{A}}_1} h \, (\partial_{\tau_2}h)^* &=& \frac{\Bar{\mathcal{A}}_2}{10 \pi } \left[ e^{i (\phi_2-\phi_1)} b_-^{(1)} b_{-,\tau_2}^{(2)}+ e^{i(\phi_1-\phi_2)} b_+^{(1)} b_{+,\tau_2}^{(2)}\right]\\
    \partial_{\phi_1} h \, (\partial_{\Bar{\mathcal{A}}_2}h)^* &=& -\frac{i \Bar{\mathcal{A}}_1}{10 \pi} \left[e^{i (\phi_2-\phi_1)} b_-^{(1)} b_-^{(2)}-e^{i(\phi_1-\phi_2)} b_+^{(1)} b_+^{(2)}\right]\\
    \partial_{\phi_1} h \, (\partial_{\phi_2}h)^* &=& \frac{\Bar{\mathcal{A}}_1 \Bar{\mathcal{A}}_2}{10 \pi} \left[e^{i (\phi_2-\phi_1)} b_-^{(1)} b_-^{(2)}+ e^{i(\phi_1-\phi_2)} b_+^{(1)} b_+^{(2)}\right]\\
    \partial_{\phi_1} h \, (\partial_{\omega_2}h)^* &=& -\frac{i \Bar{\mathcal{A}}_1 \Bar{\mathcal{A}}_2}{10\pi} \left[e^{i (\phi_2-\phi_1)} b_-^{(1)} b_{-,\omega_2}^{(2)}-e^{i(\phi_1-\phi_2)} b_+^{(1)} b_{+,\omega_2}^{(2)}\right]\\
    \partial_{\phi_1} h \, (\partial_{\tau_2}h)^* &=& -\frac{i \Bar{\mathcal{A}}_1 \Bar{\mathcal{A}}_2}{10\pi} \left[e^{i (\phi_2-\phi_1)} b_-^{(1)} b_{-,\tau_2}^{(2)}-e^{i(\phi_1-\phi_2)} b_+^{(1)} b_{+,\tau_2}^{(2)}\right]\\
    \partial_{\omega_1} h \, (\partial_{\Bar{\mathcal{A}}_2}h)^* &=& \frac{\Bar{\mathcal{A}}_1}{10\pi} \left[ e^{i (\phi_2-\phi_1)} b_{-,\omega_1}^{(1)} b_-^{(2)}+e^{i(\phi_1-\phi_2)} b_{+,\omega_1}^{(1)} b_+^{(2)}\right]\\
    \partial_{\omega_1} h \, (\partial_{\phi_2}h)^* &=& \frac{i \Bar{\mathcal{A}}_1 \Bar{\mathcal{A}}_2}{10\pi} \left[e^{i (\phi_2-\phi_1)} b_{-,\omega_1}^{(1)} b_-^{(2)}- e^{i(\phi_1-\phi_2)} b_{+,\omega_1}^{(1)} b_+^{(2)}\right]\\
    \partial_{\omega_1} h \, (\partial_{\omega_2}h)^*&=& \frac{\Bar{\mathcal{A}}_1 \Bar{\mathcal{A}}_2}{10\pi} \left[ e^{i (\phi_2-\phi_1)} b_{-,\omega_1}^{(1)} b_{-,\omega_2}^{(2)}+e^{i(\phi_1-\phi_2)} b_{+,\omega_1}^{(1)} b_{+,\omega_2}^{(2)}\right]\\
    \partial_{\omega_1} h \, (\partial_{\tau_2}h)^*&=& \frac{\Bar{\mathcal{A}}_1 \Bar{\mathcal{A}}_2}{10\pi} \left[e^{i (\phi_2-\phi_1)} b_{-,\omega_1}^{(1)} b_{-,\tau_2}^{(2)}+ e^{i(\phi_1-\phi_2)} b_{+,\omega_1}^{(1)} b_{+,\tau_2}^{(2)}\right]\\
    \partial_{\tau_1} h \, (\partial_{\Bar{\mathcal{A}}_2}h)^* &=& \frac{\Bar{\mathcal{A}}_1}{10\pi} \left[ e^{i (\phi_2-\phi_1)} b_{-,\tau_1}^{(1)} b_-^{(2)}+e^{i(\phi_1-\phi_2)} b_{+,\tau_1}^{(1)} b_+^{(2)}\right]\\
    \partial_{\tau_1} h \, (\partial_{\phi_2}h)^*&=& \frac{i \Bar{\mathcal{A}}_1 \Bar{\mathcal{A}}_2}{10 \pi } \left[ e^{i (\phi_2-\phi_1)} b_{-,\tau_1}^{(1)} b_-^{(2)}-e^{i(\phi_1-\phi_2)} b_{+,\tau_1}^{(1)} b_+^{(2)}\right]\\
    \partial_{\tau_1} h \, (\partial_{\omega_2}h)^*&=& \frac{\Bar{\mathcal{A}}_1 \Bar{\mathcal{A}}_2}{10\pi} \left[e^{i (\phi_2-\phi_1)} b_{-,\tau_1}^{(1)} b_{-,\omega_2}^{(2)} + e^{i(\phi_1-\phi_2)} b_{+,\tau_1}^{(1)} b_{+,\omega_2}^{(2)}\right]\\
    \partial_{\tau_1} h \, (\partial_{\tau_2}h)^*&=& \frac{\Bar{\mathcal{A}}_1 \Bar{\mathcal{A}}_2}{10\pi} \left[e^{i (\phi_2-\phi_1)} b_{-,\tau_1}^{(1)} b_{-,\tau_2}^{(2)}+ e^{i(\phi_1-\phi_2)} b_{+,\tau_1}^{(1)} b_{+,\tau_2}^{(2)}\right]
\end{eqnarray}
\end{widetext}
These sixteen expressions provide the FIM elements in the top right block of the total $8\times8$ matrix. To obtain the bottom left block, one can simply take the complex conjugate of the above expressions (so, for instance, $\Gamma_{\Bar{\mathcal{A}}_2 \Bar{\mathcal{A}}_1}=(\Gamma_{\Bar{\mathcal{A}}_1 \Bar{\mathcal{A}}_2})^*$).

The similarities with the expressions in Appendix~D of Ref.~\cite{Maselli:2023khq} are immediately evident. An important distinction lies in phase factors: the QNM phases are in general different for different modes, so we cannot take the exponential factors to simply be $e^0=1$.

\begin{figure*}[t]
\includegraphics[width=0.85\textwidth]{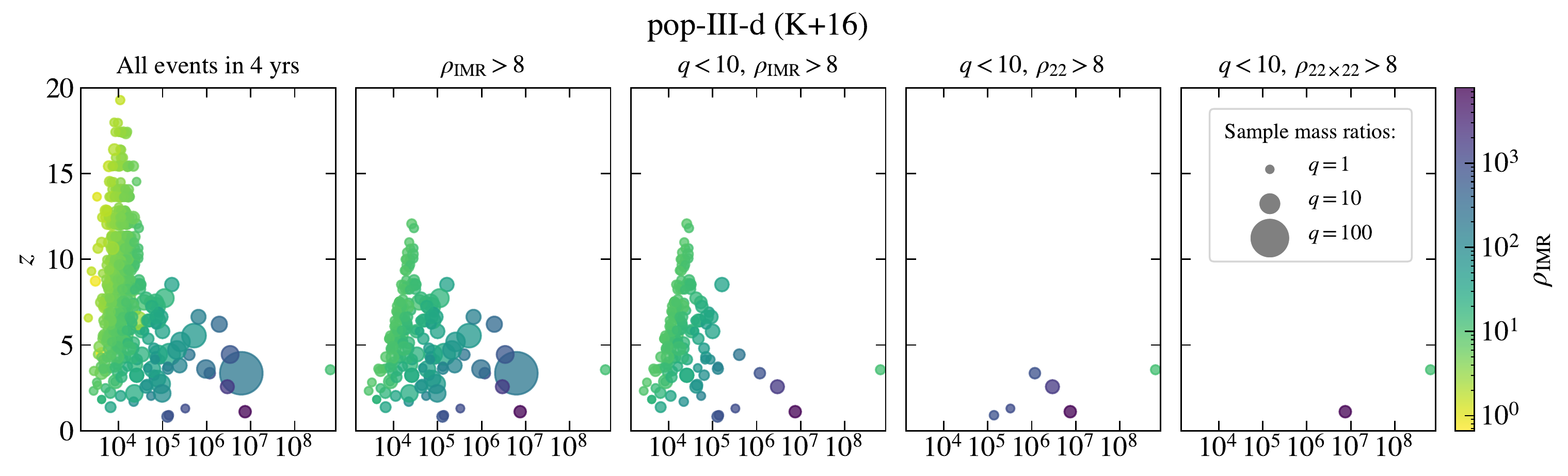}
\includegraphics[width=0.85\textwidth]{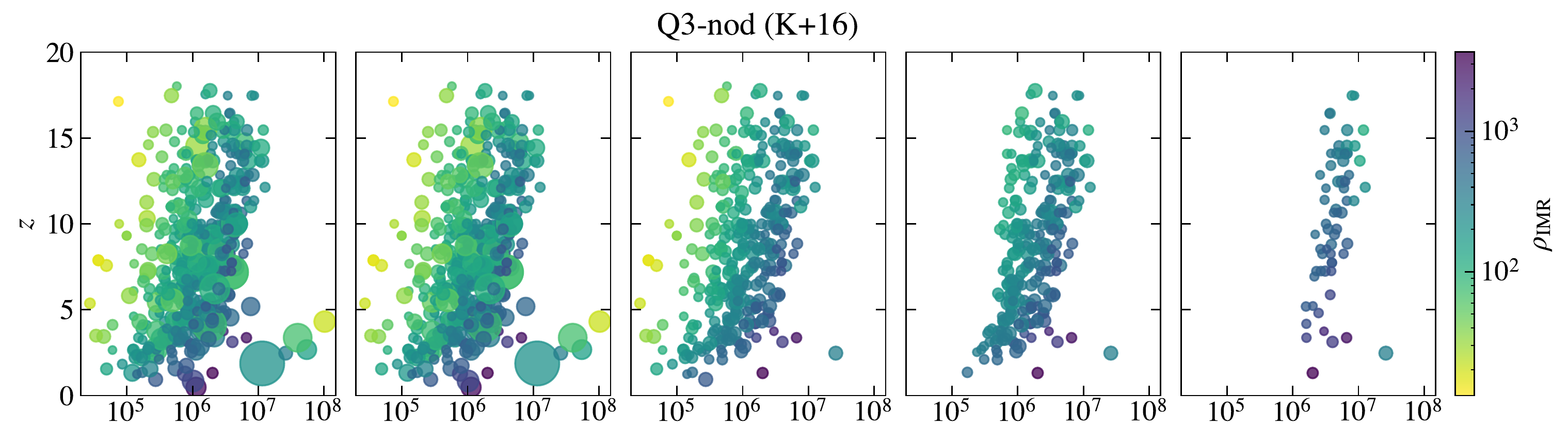}
\includegraphics[width=0.85\textwidth]{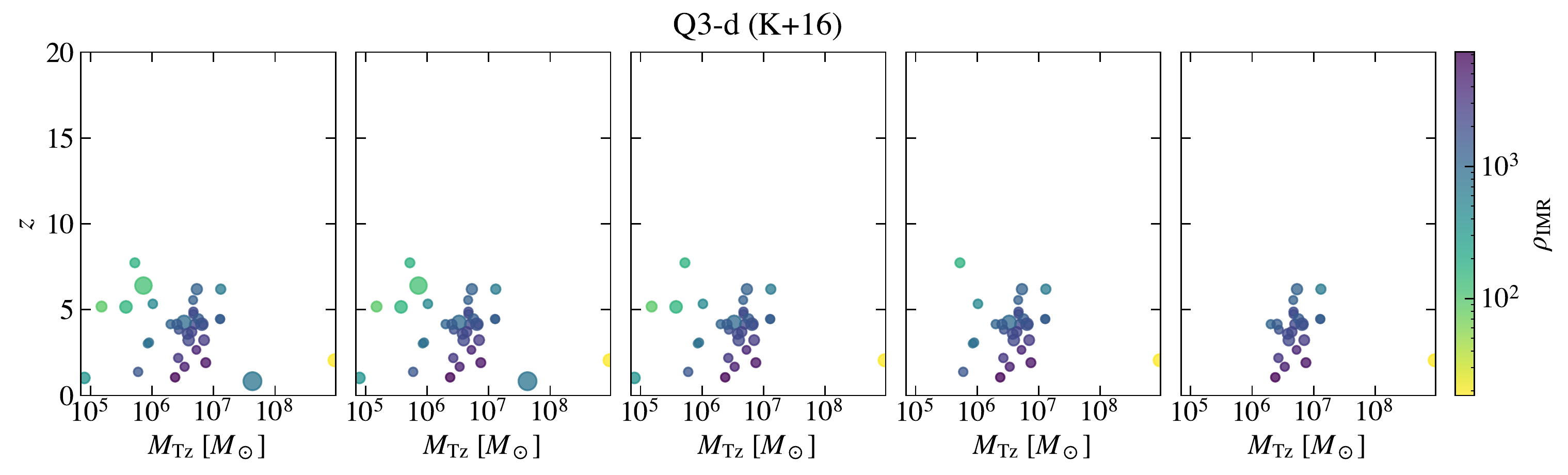}
\caption{Plots showing the relatively higher number of events with detectable ringdown and quadratic mode for heavy seed events, as compared with light seed. This figure also demonstrates that the mass ratio cut we implement only excludes a small fraction of the events in our catalogs.}
\label{fig:visualize_tbl_II}
\end{figure*}

\section{Relative number of light seed vs. heavy seed MBH events with detectable quadratic mode}\label{app:table_II_illustration}

In Fig.~\ref{fig:visualize_tbl_II}, we plot the number of MBH events satisfying various criteria in Table~\ref{tbl:MBH_stats_fin_res}. From left to right, the panels show the detector frame \emph{redshifted} total mass, $M_{\rm{T_z}}$, and the redshift, $z$, of (i) all model events in 4 years, (ii) all model events in 4 years with a detectable IMR SNR ($\rho_{\rm{IMR}}$), (iii) all model events in 4 years with detectable $\rho_{\rm{IMR}}$ and $q<10$, (iv) all model events in 4 years with detectable $\rho_{22}$ and $q<10$, and (v) all model events in 4 years with detectable $\rho_{22\times 22}$ and $q<10$. The color corresponds to the IMR SNR of the event, and the point size corresponds to $\sqrt{q}$ of the event, with a few representative mass ratio/point size correspondences given in the legend. From these plots, we see clearly that a much larger fraction of heavy seed events (``Q3-nod (K+16)'' and ``Q3-d (K+16)'') have a detectable ringdown and, in particular, a detectable quadratic mode, as compared with the light seed events (``pop-III-d (K+16)''). We also see that implementing the mass ratio cut (i.e., going from the second to the third panels in each row) in order to ensure the validity of the amplitude fits does not cause us to lose too large of a fraction of potentially detectable events.

\bibliography{nlrd}

\end{document}